\documentclass[lettersize,journal]{IEEEtran}
\usepackage{amsmath,amsfonts}
\usepackage{algorithmic}
\usepackage{algorithm}
\usepackage{array}
\usepackage[caption=false,font=normalsize,labelfont=sf,textfont=sf]{subfig}
\usepackage{textcomp}
\usepackage{stfloats}
\usepackage{url}
\usepackage{verbatim}
\usepackage{graphicx}
\usepackage{cite}
\usepackage{xurl}
\usepackage{paralist}
\usepackage{array}
\usepackage{setspace}
\usepackage{hyperref}
\usepackage{cleveref} 
\usepackage{makecell}
\usepackage{amssymb}
\usepackage{color}
\usepackage[table,dvipsnames,svgnames]{xcolor}
\usepackage{wrapfig}
\definecolor{sica}{cmyk}{0.0963, 0.0688, 0.0000, 0.1451}

\begin{document}

\title{Relation-driven Query of Multiple Time Series}

\author{
Shuhan Liu,
Yuan Tian,
Zikun Deng,
Weiwei Cui,
Haidong Zhang,
Di Weng,
Yingcai Wu, \textit{Senior Member, IEEE}

\thanks{S. Liu, Y. Tian, D. Weng, and Y. Wu are with Zhejiang University, Hangzhou, China. E-mail:
\{shliu, yuantian, dweng, ycwu\}@zju.edu.cn. D. Weng is the corresponding author.}
\thanks{Z. Deng is with the School of Software Engineering, South China University of Technology, Guangzhou, China. E-mail: zkdeng@scut.edu.cn.}
\thanks{W. Cui and H. Zhang are with Microsoft, Beijing, China. E-mail: \{weiwei.cui, haidong.zhang\}@microsoft.com.}
\thanks{Manuscript received xxx. xx, 2023; revised xxx. xx, 2023.}
}

\markboth{Journal of \LaTeX\ Class Files,~Vol.~14, No.~8, August~2021}%
{Shell \MakeLowercase{\textit{et al.}}: A Sample Article Using IEEEtran.cls for IEEE Journals}


\maketitle

\begin{abstract}
Querying time series based on their relations is a crucial part of multiple time series analysis.
By retrieving and understanding time series relations, analysts can easily detect anomalies and validate hypotheses in complex time series datasets.
However, current relation extraction approaches, including knowledge- and data-driven ones, tend to be laborious and do not support heterogeneous relations.
By conducting a formative study with 11 experts, we concluded six time series relations, including correlation, causality, similarity, lag, arithmetic, and meta, and summarized three pain points in querying time series involving these relations.
We proposed RelaQ, an interactive system that supports the time series query via relation specifications.
RelaQ allows users to intuitively specify heterogeneous relations when querying multiple time series, understand the query results based on a scalable, multi-level visualization, and explore possible relations beyond the existing queries.
RelaQ is evaluated with two cases and a user study with 12 participants, showing promising effectiveness and usability.

\end{abstract}


\section{Introcution}
\begin{figure}[htbp]
  \includegraphics[width=\linewidth]{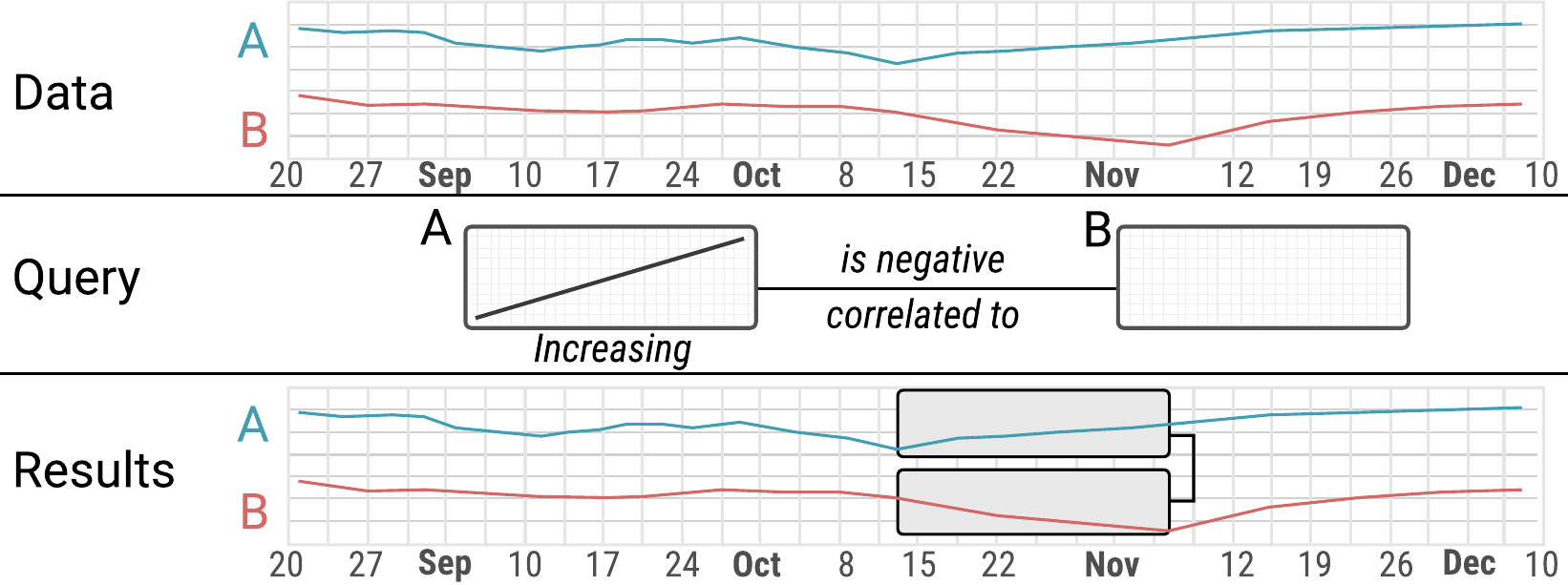}
  \caption{An example of a relation-driven query. An analyst observed a global positive-correlated pair of time series (A and B) and sought local negative-correlated fragments. Finally, the result time fragments were highlighted.}
  \label{fig:stock}
\end{figure}

Many research efforts \cite{peax, PSEUDo, kdbox} have been devoted to the interactive query of multiple time series, empowering users to find patterns from large-scale time series data quickly.
Among the constraints used in such queries, time series relations, {such as correlation~\cite{cpc, hongcorr}, causality~\cite{domino, MECau}, and similarity~\cite{comparesim, PSEUDo}}, are frequently employed to describe the patterns that span across multiple time series.
For instance, when analyzing a stock dataset with multiple time series, an analyst may query two time fragments (e.g., one month) that have a local negative correlation while the two time series have a positive correlation overall (e.g., one year), as shown in the Fig.~\ref{fig:stock}

Most of the current approaches for the interactive query of multiple time series identify the desired time series fragments based on the given thresholds~\cite{timesearcher2005,ivquery} and trend patterns~\cite{SDL,querysketch,lstm2021}, such as the \textit{head and shoulders} patterns~\cite{qetch}.
Nonetheless, while these approaches primarily measure the similarity between query specifications and the resulting time series, they unfortunately do not facilitate the specification of relationships amongst these time series.
Due to the lack of interactive approaches for relation-driven time series queries, analysts often resort to writing lengthy scripts with statistical libraries (e.g., Pandas~\cite{Pandas}) or tools (e.g., tradingview~\cite{tradingview}), which can be laborious and error-prone.
Moreover, analyzing the relations based on the queried time series fragments is also a difficult task: \textit{How do these fragments distribute? Are the strengths of these relations consistent over time? Are there more relations between these fragments and the rest of the data?}

To better understand the practice, challenges, and requirements of the relation-driven query of multiple time series, we conducted a formative study with eleven time-series analysts from different domains.
This study helps us summarize a) in \textit{what} time series relations the analysts are interested; b) \textit{why} these relations are employed in multiple time series analysis; and c) \textit{how} these relations {are queried} and support the workflows of the analysts.
We conclude the following three challenges in designing a new relation-driven time series query approach from our observations.

\textbf{Diverse and heterogeneous time series relations.}
In the formative study, we have identified six types of time series relations.
Analysts must retrieve the multiple time series satisfying multiple types of relation constraints.
Not only are new interactions required to enable the easy specification of the relations among time series, but a new retrieval algorithm is also needed to efficiently find the matches satisfying the different relations in one query.

\textbf{Intuitive interpretation of complex query results.}
The results of relation-driven time series queries can be difficult to comprehend, given that many matches can be produced from a large-scale time series dataset.
For example, analysts need to check the fluctuation in the strengths of the queried relations to determine whether the relations become more consistent over time.
Moreover, the incorporation of relaxed retrieval, which allows partial matching to increase the diversity of query results, also introduces the difficulty in understanding the structural changes in the resulting time series relations.

\textbf{Efficient generation of reliable relation suggestions.}
Inspired by the query suggestions provided by modern search engines, the proposed approach should support the exploration of the potential relations among the time series of interest beyond what have been specified in the queries.
However, given the sheer volume of time series and the diverse types of time series relations, it is challenging to efficiently generate and provide useful relation suggestions to complement the existing queries.

\begin{figure*}[t]
    \centering
\includegraphics[width=\linewidth]{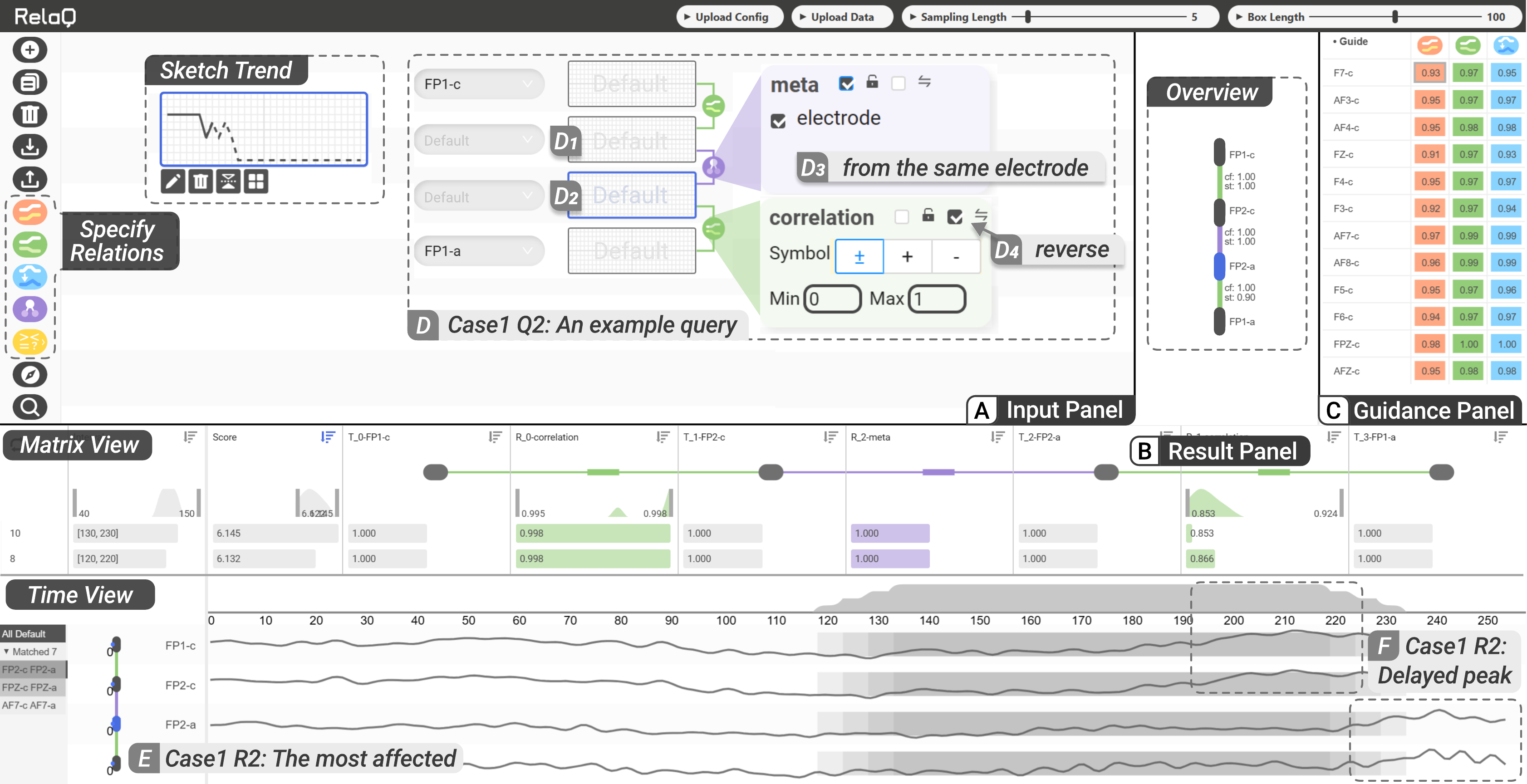}
\caption{The user interface of RelaQ. It is composed of three main parts: In the (A) \textit{input} panel, users can sketch trends and specify relations. In the (B) \textit{result} panel, users can obtain an overview, compare results in the matrix view, and inspect details in the time view. In the (C) \textit{guidance} panel, there are recommended timeboxes. Besides, this figure also displays part of the first case in Sec.~\ref{sec:case1}: (D) an example query and (E-F) some patterns in its results.}
\label{fig1.teaser}
\end{figure*}

We propose RelaQ, an interactive approach that retrieves multiple time series based on their relations.
To address the above challenges, RelaQ comprises a query interface (Fig.~\ref{fig1.teaser}) that allows users to visually specify various relations among time series and processes the queries with a new search algorithm that supports matching time series with heterogeneous relation constraints.
The query interface is composed of three parts (Fig.~\ref{fig1.teaser}A-C): an input panel, a result panel, and a guidance panel.
A scalable visual design of the result panel, depicting the topologies and characteristics of the matched results, is incorporated to visualize the resulting time series at different levels of detail, allowing users to perform multiscale analysis.
Moreover, the guidance panel displays an overview of the suggestions that extend the existing queries with more relations and time series.

Our contributions are summarized as follows:
\begin{compactitem}
    \item A formative study that discusses the scope of time series relations and their applications and summarizes the challenges and requirements in the relation-driven query of multiple time series;
    \item A novel approach that combines a fuzzy query model and an interactive interface to support the easy formulation of heterogeneous relation constraints, the {flexible} query of multiple time series based on specified relations, and the intuitive interpretation of the query results.
\end{compactitem}

\section{Related Work}
\label{sec:related_work}
We mainly reviewed time series studies on query specification, query matching, and query results visualization.
\subsection{Time Series Query Specification}
Specifying a query is the first step of time series retrieval.
Intuitive specification approaches facilitate users to formulate queries efficiently.
According to different input forms, existing methods can be divided into four categories: text-, value-, example-, and sketch-based query.

The \textbf{text-based} query allows expressing desired patterns through natural language or regex.
Agrawal et al.~\cite{SDL} were the first to propose a shape definition language (SDL) that describes time series patterns.
SDL assisted in formulating a time series into trend shape symbol sequences (e.g., up, down, and stable).
In recent years, there have also been many methods to support regex~\cite{shapesearch} and constrained natural languages~\cite{qute}.
However, text-based methods suffer from the concern of learnability and intuitiveness because users must convert imagined patterns into a certain abstract language.

The \textbf{value-based} query requires to set value bounds to query time series passing through.
TimeSearcher~\cite{timesearcher2002,timesearcher2005} was the first to adopt timeboxes to identify value constraints.
A timebox is a rectangle, where the horizontal axis represents the length constraints of a time range, while the vertical axis represents the value bounds of time series data.

Later, the \textbf{example-based} methods extensively consider time series within a timebox as examples for querying similar patterns.
Many models built based on examples make the real-time performance of querying time series significantly improved, such as PSEUDo~\cite{PSEUDo}, PEAX~\cite{peax}, and KD-Box~\cite{kdbox}.
However, value- and example-based methods are not flexible to describe various trend shapes like \textit{``head-and-shoulder''} {\def\svgwidth{0.5cm}%% Creator: Inkscape 1.3 (0e150ed, 2023-07-21), www.inkscape.org
%% Accompanies image file 'HAS.pdf' (pdf, eps, ps)
\begingroup%
  \makeatletter%
  \providecommand\color[2][]{%
    \errmessage{(Inkscape) Color is used for the text in Inkscape, but the package 'color.sty' is not loaded}%
    \renewcommand\color[2][]{}%
  }%
  \providecommand\transparent[1]{%
    \errmessage{(Inkscape) Transparency is used (non-zero) for the text in Inkscape, but the package 'transparent.sty' is not loaded}%
    \renewcommand\transparent[1]{}%
  }%
  \providecommand\rotatebox[2]{#2}%
  \newcommand*\fsize{\dimexpr\f@size pt\relax}%
  \newcommand*\lineheight[1]{\fontsize{\fsize}{#1\fsize}\selectfont}%
  \ifx\svgwidth\undefined%
    \setlength{\unitlength}{108.87293297bp}%
    \ifx\svgscale\undefined%
      \relax%
    \else%
      \setlength{\unitlength}{\unitlength * \real{\svgscale}}%
    \fi%
  \else%
    \setlength{\unitlength}{\svgwidth}%
  \fi%
  \global\let\svgwidth\undefined%
  \global\let\svgscale\undefined%
  \makeatother%
  \begin{picture}(1,0.65121677)%
    \lineheight{1}%
    \setlength\tabcolsep{0pt}%
    \put(0,0){\includegraphics[width=\unitlength,page=1]{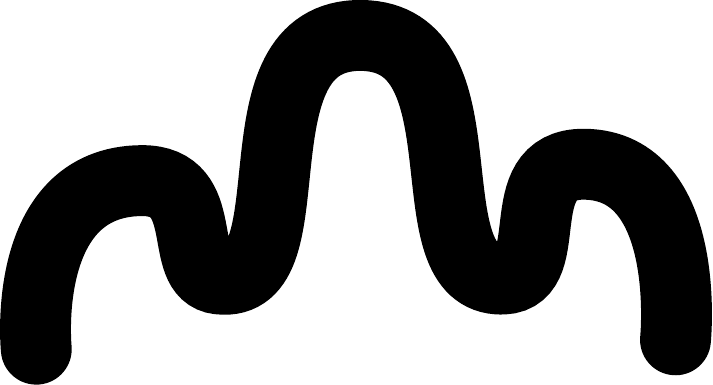}}%
    \put(0.15202411,0.02876841){\color[rgb]{0,0,0}\makebox(0,0)[lt]{\lineheight{1.25}\smash{\begin{tabular}[t]{l}\end{tabular}}}}%
  \end{picture}%
\endgroup%
}, especially when the desired example is hard to locate.

In the \textbf{sketch-based} query methods, users can sketch desired trend shapes to retrieve similar time series.
We followed Mannino et al.~\cite{qetch} and divided existing sketch-based approaches into three categories: overlays, annotated sketches, and shape-restricted sketches.
Overlaying approaches~\cite{querysketch,lstm2021,semanticsketch} leveraged the predefined time and amplitude axes to make the scale clear but were highly dependent on the selected reference time series.
Annotated sketches approaches (e.g., Qetch~\cite{qetch}, Holz et al.~\cite{holz2009circle}) allowed users to freely hand-draw desired patterns on blank canvas but required providing scale or pattern descriptions. 
Many visual query systems allowed users not to provide scale annotations but to adaptively match patterns based on restricted shapes ~\cite{Keogh1997straight,shapegrammar}.
These visual query systems had good robustness, but slacked scale constraints led to a high recall rate.
While sketching is useful and expressive, none of the existing methods supports specifying multiple time series and their relations.
Users cannot always sketch what they want, such as two correlated time series.

In addition, existing methods prioritize similarity between the input query and output results but ignore relations among different time series in one query.
Thus, we are inspired to propose RelaQ, which supports querying multiple time series with heterogeneous relations.
RelaQ is based on sketching because it describes temporal features expressively.

\subsection{Time Series Query Matching}
Most \textbf{general time series query matching} approaches focus on retrieving similar time series, so many methods have been proposed to evaluate the similarity between the query and the results.
Euclidean distance (ED)~\cite{ed} and dynamic time warping (DTW)~\cite{dtw} are popular for measuring distances between time series.
According to the survey of Ding et al.~\cite{ding2008survey}, DTW is the best among all distance measures, but ED is faster so can be used when the scale of data increases.
Both ED and DTW require a sliding window technique.
Some researchers also transform time series into symbolic sequences and exploit string-matching techniques.
Symbolic Aggregate approXimation (SAX)~\cite{sax} is a proven efficient tool.
SAX compresses the length by segmenting the time series and reduces the cardinality by discretizing and symbolizing the values.
There are also many approaches based on machine learning techniques~\cite{lstm2021, dr} or designed for special situations~\cite{kdbox}.
For large-scale time series data, there are mature commercial database query tools available, including time series databases (e.g., InfluxDB~\cite{InfluxDB}, Timestream~\cite{Timestream}), relational time series databases (e.g., TimescaleDB~\cite{TimescaleDB}), and general search databases (e.g., Elasticsearch~\cite{ElasticSearch}).

As for \textbf{relation-driven query matching}, despite many studies devoted to helping query, extract, and analyze relations between time series, most are designed for certain relations, such as correlation~\cite{irvine,ecoalvis}, causality~\cite{compass,jin2021causal}, co-occurrence~\cite{LeeTrend,SHIRATO202357,SHIRATO202377}).
In other words, algorithms and tools that support matching heterogeneous relations have still not been sufficiently studied.
If users want to obtain complex relations in real analysis scenarios, they can only return to scripts or commercial tools, such as tradingview~\cite{tradingview}. % EEGLAB

Thus, we propose a flexible algorithm that simultaneously matches trends and heterogeneous relations, enabling RelaQ to support complex relation-driven queries.

\subsection{Dynamic Graph Visualization}
Intuitively visualizing query results is beneficial for users to refine queries or perform further exploration.
Query results are dynamic graphs of time fragments (nodes) with changing relations (links), like two stocks that may shift from positive to negative correlation and become unrelated over time.
Beck et al.~\cite{graphsurvey} developed a three-level taxonomy for dynamic graph visualization techniques.
The first level includes three types: animation, timeline, and hybrid.

As relations are integrated closely with time lags and trends, a timeline-based visualization could enable time-oriented exploration of query results.
Hence, we mainly reviewed linear timeline techniques~\cite{mickschdynamic}.
Burch et al.~\cite{Burch13matrix} embedded a timeline into each cell of the matrix design, while Bach et al.~\cite{cubix} stacked matrices to a 3D cube by timeline.
Besides, Hlawatsch et al.~\cite{Hlawatsch14list} introduced a visualization based on adjacent lists.
However, neither matrix-based nor list-based design is intuitive for depicting time lags.
The node-link-based designs mapped time to space and created lag-aware static images, which assist users in understanding the temporal graph evolution.
Beck et al.~\cite{graphsurvey} categorized these techniques into juxtaposed \cite{BurchVBDW11,timearctree}, superimposed \cite{BrandesC03}, or integrated \cite{ShiWWQLL15} layouts.

Moreover, since the node-link-based visualization techniques underperform on the comparison task, we consider integrating matrix-based methods.
Inspired by LineUp~\cite{Lineup}, we design an exceptional matrix-node-integrated visualization to balance the data nature and user experience.
It also supports flexible decision-making on dynamic temporal graphs.
\section{Formative Study}
\label{sec:formative}
This section presents a formative study that aims to discover the pain points and needs in the time series query from the relation perspective and further compile the user requirements that guide the development of a relation-driven time series retrieval tool.
By interviewing time series analysts, we attempted to gather insights on three research questions: a) \textit{what} time series relations were considered important to their daily analysis tasks; b) \textit{why} their analysis tasks required these time series relations; and c) \textit{how} they {retrieved and} leveraged these time series relations to support their analysis tasks.

\subsection{Method}

To answer the above questions, we first conducted a literature survey to search for the types of time series analysis tasks and narrow down the scope of time series relations.
Then, we prepared seven interview questions and conducted semi-structured interviews individually with eleven time series analysts.
Finally, we analyzed their responses with the thematic coding approach~\cite{gibbs2007thematic} and compiled three major user requirements that supported the design of the proposed tool.

\textbf{Literature survey.}
To produce an initial list of time series relations, we searched for the papers published in the prominent data mining, visualization, and human-computer interaction conferences and journals that contained keywords including \textit{multiple time series} and \textit{time series relation}.
The surveyed conferences and journals include but are not limited to ACM KDD, IEEE VIS, IEEE TVCG, ACM CHI, etc.
Two co-authors with years of experience in time series analysis went through 112 papers separately and tagged the keywords that described the time series relations used in the papers.
After reconciling and merging the keywords, we summarized ten relations: \textit{similarity}, \textit{correlation}, \textit{causality}, \textit{co-occurrence}, \textit{precedence}, \textit{cascade}, \textit{lag}, \textit{cycle}, \textit{hierarchy}, and \textit{group}.

\textbf{Semi-structured interviews.}
To refine the preliminary scope of time series relations and understand how these relations were involved in the time series analysis tasks, we recruited eleven time series analysts (five male and six female) as participants and conducted semi-structured interviews.
The participants had diverse backgrounds such as energy (1), robot simulation (2), urban computing (1), stock trading (2), cloud services (3), and sports computing (2), and each participant had a minimum of two years of experience in analyzing time series data.
They did not receive monetary compensation for participating in this formative study.
The interviews were conducted via video telephony due to pandemic restrictions and followed the questions listed below (\texttt{M} denotes multiple-choice questions, \texttt{S} denotes single-choice questions, and \texttt{O} denotes open-ended questions):

\begin{compactitem}
    \item[\textbf{Q1.}] \texttt{[M]} Which time series analysis tasks do you usually work on?
    \item[\textbf{Q2.}] \texttt{[S]} How many times per week do you work on time series tasks?
    \item[\textbf{Q3.}] \texttt{[O]} What is the scale of the time series data (in the order of magnitude) you are dealing with?
    \item[\textbf{Q4.}] \texttt{[M, O]} Which relations do you want to retrieve during the analysis process?
    \item[\textbf{Q5.}] \texttt{[O]} Which insights do you gain from these relations, and how important are they?
    \item[\textbf{Q6.}] \texttt{[O]} How do you typically retrieve these time series relations in your data? Is there any limitation?
    \item[\textbf{Q7.}] \texttt{[O]} In existing approaches, are there any advantages or functions you prefer? After retrieving time series relations, what else will you explore?
\end{compactitem}

Q1-3 were designed to learn the participants' expertise and routines, including their familiar analysis tasks (Q1), experience profile (Q2), and data scale (Q3).
Q1 has 9 choices (Fig.~\ref{fig2.experts}A) based on the prior surveys of time series analysis tasks~\cite{Fakhrazari17survey,timeseriessurvey12}.
For Q3, we asked the participants to estimate the average length and the number of time series.
Q4-7 were designed to solicit the participants' opinions on the scope of time series relations, including the desired relations in time series query (Q4), the usages and importance of the relations (Q5), the prior approaches used in the query (Q6), and analysis preferences and requirements (Q7).
Q4 is a multiple-choice and open-ended question, where we first asked the participants to choose from our initial list of time series relations and then asked the participants if some relations should be removed and/or more relations should be added.
Each interview lasted between 30-50 minutes.
The participants' responses were recorded and coded with the thematic coding approach~\cite{gibbs2007thematic}.

\subsection{Results}

\subsubsection{Participants' Backgrounds (Q1-3)}
\begin{figure}[t]
  \centering
  \includegraphics[width=\linewidth]{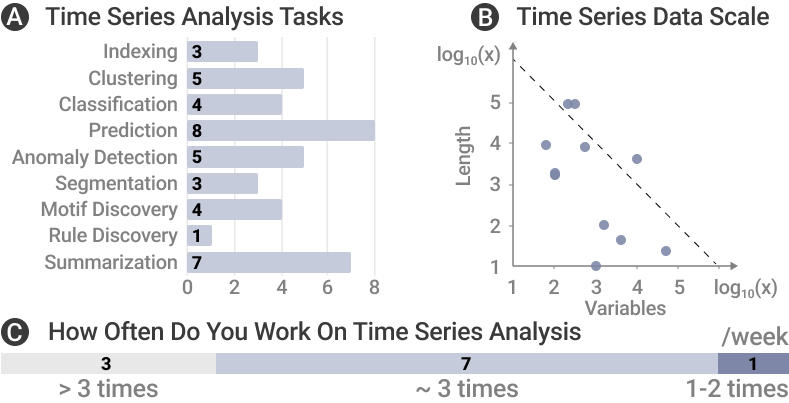}
  \caption{
     Investigation responses from time series analysts.
     (A) Typical time series analysis tasks that participants work on, all nine types of tasks are covered. (B) The scale of time series data that participants usually deal with, the number of \textbf{variables} $\times$ the \textbf{length} of a single time series. (C) Frequency that analysts analyze time series in a week, reflecting all participants are very familiar with time series data.
  }
  \label{fig2.experts}
\end{figure}

Fig.~\ref{fig2.experts}A displays experts' familiar time series analysis tasks (Q1).
Most of the participants are familiar with more than two types of analysis tasks, and prediction and summarization are the most common.
Though only P5 has worked on rule discovery, all other analysis tasks are mentioned by at least three participants.

We investigate participants' experience profiles via their working frequency (Q2).
Fig.~\ref{fig2.experts}C reveals that most participants maintained a frequency of three or more times a week.

To paint a clearer picture of the participants' analysis work, we survey the scale of the time series data (Q3).
Since multiple time series have two dimensions (the number of series and the length of each time series), we asked participants to describe both.
According to Fig.~\ref{fig2.experts}B, the number of time series ranges in order of magnitude from $10$ to $10^4$, while the lengths range in order of magnitude from $10$ to $10^5$.

Thus, the diverse backgrounds and high average proficiency of our participants make us confident that various domains and tasks are covered in our formative research.
\subsubsection{Time Series relations Scope (Q4-5)}
\label{sec:scope}

We summarized six categories of relations, including \textit{similarity}, \textit{correlation}, \textit{causality}, \textit{lag}, \textit{meta}, and \textit{arithmetic}, based on the participants' responses.
\textit{Co-occurrence}, \textit{precedence} and \textit{cascade} were removed because they can be represented by the combination of \textit{lag} and other relations; \textit{cycle} was removed because it mainly focuses on the temporal characteristics of individual time series;
and \textit{hierarchy} and \textit{group} were aggregated into \textit{meta}, which captures the latent semantic relations between time series.
Besides, we added \textit{arithmetic} according to participants' analysis requirements. 

We started the introduction of the scope from time series' different granularity, from high to low: time \textbf{S}eries, \textbf{F}ragments, and \textbf{P}oints.
The strengths of a relation vary on different granularity (see TABLE~\ref{tab1.scope}).
This section will discuss the definition, {usage}, and calculation of time series relations.
\begin{table}[t]
\caption{The granularity and adopted calculation methods of the time series relations. All relation strengths are normalized.}
\centering
\renewcommand{\arraystretch}{1.25}
 \setlength{\tabcolsep}{1.5mm}{
 \footnotesize 
\begin{tabular}{|l|l|l|l|l|l|} 

\hline
\rowcolor{lightgray!50}
\multicolumn{1}{|l|}{\sf  {\textbf{Relations}}}
& \multicolumn{3}{l|}{\sf  \textbf{Granularity}}
& \multicolumn{2}{l|}{\sf  \textbf{Calculation Methods}}  \\ 
\hline

\rowcolor{sica!75}
{\sf  Names}
&  {\sf  P} & {\sf  F} & {\sf  S}
&  {\sf  RelaQ Adopted} & {\sf  Domains} \\ 

\hline
{\sf  Correlation} & 
\makecell[c]{\sf $\times$}
&\makecell[c]{\sf \checkmark}&\makecell[c]{\sf \checkmark}& {\sf  Pearson Coefficient~\cite{pearson1901liii}} & {\sf  [-1.0,+1.0]} \\ 

\hline
{\sf  Similarity} & \makecell[c]{\sf \checkmark} &\makecell[c]{\sf \checkmark}&\makecell[c]{\sf \checkmark}& {\sf  Euclidean Distance~\cite{ed}} &\makecell[l]{\sf [+0.0,+1.0]}\\         

\hline
{\sf  Causality}& \makecell[c]{\sf \checkmark} &\makecell[c]{\sf \checkmark}&\makecell[c]{\sf \checkmark}& {\sf  Granger Test~\cite{granger1969investigating}} & {\sf  [+0.0,+1.0]} \\ 

\hline
{\sf  Lag} & \makecell[c]{\sf \checkmark} &\makecell[c]{\sf \checkmark}&\makecell[c]{\sf \checkmark} & \multicolumn{2}{l|}{\sf  Attached to other relations} \\

\hline
{\sf  Meta} & \multicolumn{3}{c|}{\sf \checkmark} & {\sf  Predefined by users} & {\sf  {0-N, 1-Y}}\\ 

\hline
{\sf  Arithmetic} & \makecell[c]{\sf \checkmark} &\makecell[c]{\sf \checkmark}&\makecell[c]{\sf \checkmark}& \multicolumn{2}{l|}{\sf  $\{\sum, Avg, Var, Min, Max\}\times\{\geq, \leq, =$\}}\\ 
\hline
\end{tabular}
}
\label{tab1.scope}
\end{table}

\begin{wrapfigure}[3]{l}{0.7cm}
% \centering
\vskip-\intextsep
\includegraphics[width=0.9cm]{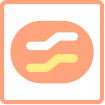}
\vskip-\intextsep
\end{wrapfigure}
\noindent\textbf{Similarity:}
Similarity between two time series refers to the degree to which they exhibit comparable patterns and behavior.
The similarity relation is widely used in many analysis tasks, especially those that depend on extracting commonalities or differences.
For tasks including clustering (P1, P3, P6, P8, P10), indexing (P2, P4), and prediction (P1, P11), time series with high similarity are important.
For similarity-aware anomaly detection (P4-5, P8), participants are interested in time series with low similarity.
According to the interviews, Euclidean distance~\cite{ed}, dynamic time warping~\cite{dtw}, and MAPE~\cite{MAPENew} are commonly used similarity measurements.

\begin{wrapfigure}[3]{l}{0.7cm}
% \centering
\vskip-\intextsep
\includegraphics[width=0.9cm]{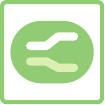}
\vskip-\intextsep
\end{wrapfigure}
\noindent\textbf{Correlation:}
Correlation between two time series refers to the degree to which they exhibit {an associated} relation with each other.
Correlation is one of the most important relations serving various time series analysis tasks, such as clustering (P1, P3, P6), classification (P6-7), prediction (P3, P6), anomaly detection (P6), and motif/rules discovery (P4-5).
There are two categories of correlations: linear and non-linear.
The most commonly used way to calculate the linear correlation is Pearson coefficient~\cite{pearson1901liii}.
Also, there are approaches for non-linear correlation (SROCC~\cite{spearman1906footrule} and Copula~\cite{li2000default}) and specific domain measures (P7:PLV~\cite{lachaux1999measuring}).

\begin{wrapfigure}[3]{l}{0.7cm}
% \centering
\vskip-\intextsep
\includegraphics[width=0.9cm]{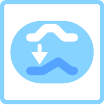}
\vskip-\intextsep
\end{wrapfigure}
\noindent\textbf{Causality:}
Causality between two time series refers to the relation where one time series, known as the cause, has a direct influence on the other time series, known as the effect.
The causality is critical in the prediction task (P1, P5-6, P11).
Participants often query time series with causal relations in history to validate predicted results.
Besides, it is sometimes adopted to help mine motifs or rules as well as detect anomalies (P5-6).
The causality is usually established by causal inference techniques, such as Granger causality~\cite{granger1969investigating} or Bayesian networks~\cite{bayesian}.
The strength can be computed via p-value and Bayesian probability, respectively.

\begin{wrapfigure}[3]{l}{0.7cm}
% \centering
\vskip-\intextsep
\includegraphics[width=0.9cm]{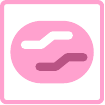}
\vskip-\intextsep
\end{wrapfigure}
\noindent\textbf{Lag:}
Lag relation between two time series refers to the temporal delay or phase shift between them.
The lag relation reflects the time dimension independently and can be combined with other five relations.
Participants usually query lag relation in analysis tasks to obtain temporal features, for example, prediction (P3, P6-7, P9-10) and periodical motif discovery (P4, P7).
The time lag is usually calculated by cross-correlation~\cite{1927cross}, self-correlation (P7), n-derivation (P3), or inferred from domain knowledge (P7, P9-10).

\begin{wrapfigure}[3]{l}{0.7cm}
% \centering
\vskip-\intextsep
\includegraphics[width=0.9cm]{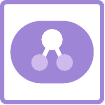}
\vskip-\intextsep
\end{wrapfigure}
\noindent\textbf{Meta:}
The meta relation between two time series refers to the semantic or contextual attributes that they share, usually about hierarchies and categories.
For example, two time series representing daily births in LA and California state may have a meta relation due to their hierarchical structure.
The meta relation provides contextual information and helps understand time series in context.
Participants usually adopt it as a kind of filter.
It is widely used in time series analysis tasks, such as clustering (P1, P3), classification (P3), and semantic anomaly detection (P8).
Participants usually measure it through predefined data configuration.

\begin{wrapfigure}[3]{l}{0.7cm}
% \centering
\vskip-\intextsep
\includegraphics[width=0.9cm]{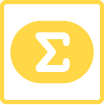}
\vskip-\intextsep
\end{wrapfigure}
\noindent\textbf{Arithmetic:}
Arithmetic relation between two time series refers to the statistical features.
Specifically, the arithmetic relation is usually composed of an operator and a comparator.
Participants usually query arithmetic relations to locate time series with specific statistical features.
For example, the average (operator) value of time series A is greater (comparator) than that of B.
According to the interview, participants query for arithmetic relations in many analysis tasks: classification (P7), prediction (P9), and anomaly detection (P8, P10).
Commonly used operators and comparators are listed in TABLE~\ref{tab1.scope}.
\subsubsection{User Requirements Analysis (Q6-7)}
\label{sec:requirements}
Based on the participants' responses, we summarized two types of queries (target- and breadth-oriented) and three user requirements.

\noindent\textbf{R1. Straightforward query of desired time series.}
According to the participants' responses, 
a straightforward approach is required to support intuitive query formulation and {flexible} query processing since there were many limitations in their daily-used approaches.

In terms of query formulation, most participants found describing the relation among time series to be the most challenging part.
Currently, the method for specifying relations involves writing scripts and rules, which participants described as \textit{``an empirical work''} and a \textit{``heavy mental burden''}.
They must transform abstract relations into concrete rules (e.g., if-else statements, P2, P7, P11) or accurate parameters (P3, P4-6, P8).
Many participants struggled with this transformation and expressed a desire to describe relations directly.
Additionally, participants discussed constraints, such as values and trends, that describe the inner features of time series. While many participants agreed that sketching trends to query time series was expressive (P1-2, P5-7, P11), they found it challenging to adopt this approach in their analysis workflow.
P5 explained that she often needed to search for correlated time series with partly unknown trends, which she could not sketch entirely.
Nevertheless, participants still found sketching trends to be an intuitive approach and expressed a desire to use it.

In terms of query processing, participants expressed concern about processing strategy, highlighting the importance of flexible processing in reducing tedious query constraint refinement.
They were not always confident that their queries would be entirely correct on the first attempt (P2, P5).
Some participants suggested a fuzzy-match approach to processing queries, as it could include more potential results (P1, P3).

\noindent\textbf{R2. In-depth understanding of query results.}
\label{sec:R2}
Participants reported that the in-depth understanding of query results primarily involves evaluating and identifying queried patterns within time series.
Thus, a comprehensive and interactive display of results is essential for supporting such a requirement.

Nearly all participants expressed a desire to prioritize query results based on diverse metrics, such as confidence (P8), similarity (P9-11), correlation (P1, P11), and matching score (P3, P7).
However, they often had to manually calculate and balance between multiple metrics.
P1, a stock analyst, reported that homogeneous evaluation criteria can affect feature selection and thus prediction results, prompting a desire to identify evaluation metrics dynamically.
Overall, an interactive approach is needed to support flexible evaluation, allowing customizing metrics based on specific needs.

Moreover, some participants found it cumbersome to locate desired patterns accurately in query results.
For example, P9, a software analyst, told us that she usually visualized results manually using matplotlib, then inspected the beginning timestamp of anomaly increasing by naked eyes.
To overcome the limitations of such a tedious and inaccurate inspection process, an intuitive display of query results is needed to support the analysis of detailed time series and relations.

\noindent\textbf{R3. Exploration based on reliable guidance.}
The majority of participants reported that the current query process was time-consuming.
Upon further investigation into participants' workflow, we identified the most tedious stage: exploring desired queries.
This often occurs when analysts did not have a specific target in mind, and therefore needed to explore desired queries by iteratively finding related relations and extending existing queries.
It differs from the situations previously discussed in \textbf{R1}: participants could describe a clear target, such as querying two correlated stocks.
Taking inspiration from Heer et al.~\cite{voyager}, we categorized these two types of queries as breadth-oriented and target-oriented, respectively, with participants needing to perform both types for complex time series analysis tasks.

Breath-oriented queries required participants to find new related relations and extend existing queries in large-scale time series data. Due to complexity, participants reported getting lost in the exploration process (P2).
Therefore, many participants (P1-2, P4, P6, P11) expected reliable guidance.
Specifically, step-by-step guidance is preferred (P2, P4) as it gives them full control over the query.
\section{RelaQ}
\label{sec:system}
\begin{figure}[b]
  \centering
  \includegraphics[width=\linewidth]{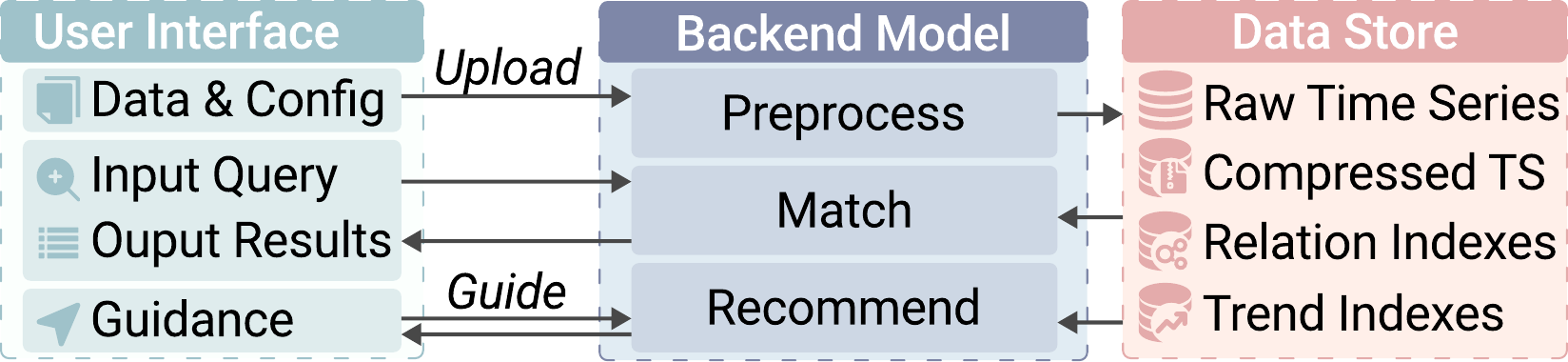}
  \caption{
     The overview of RelaQ's workflow.
     Users upload a multiple time series dataset and a configuration file to RelaQ, which preprocesses the data, builds various indexes, and allows users to specify queries. The system can also provide guidance by recommending query constraints.
  }
  \label{fig3.overview}
\end{figure}
This section presents RelaQ, an interactive time series retrieval tool based on relation queries, which was carefully designed to support the requirements summarized in Sec.~\ref{sec:requirements}.
The workflow of using RelaQ is illustrated in Fig.~\ref{fig3.overview}.
First, users can upload a dataset consisting of multiple time series, along with a configuration file describing the semantic labels (e.g., group of belonging) for each time series.
RelaQ preprocesses the uploaded data into compressed time series, relation indexes, and trend indexes.
Second, users can specify query constraints among interested time series with the input panel, and RelaQ finds and returns matches for the input query with a search model.
Third, RelaQ can provide guidance in query specification by offering additional time series relations within the dataset.
RelaQ is implemented based on React-Redux-TS~\cite{rrt} and Flask-Python~\cite{flask}.

\begin{figure}[hb]
  \centering
  \includegraphics[width=\linewidth]{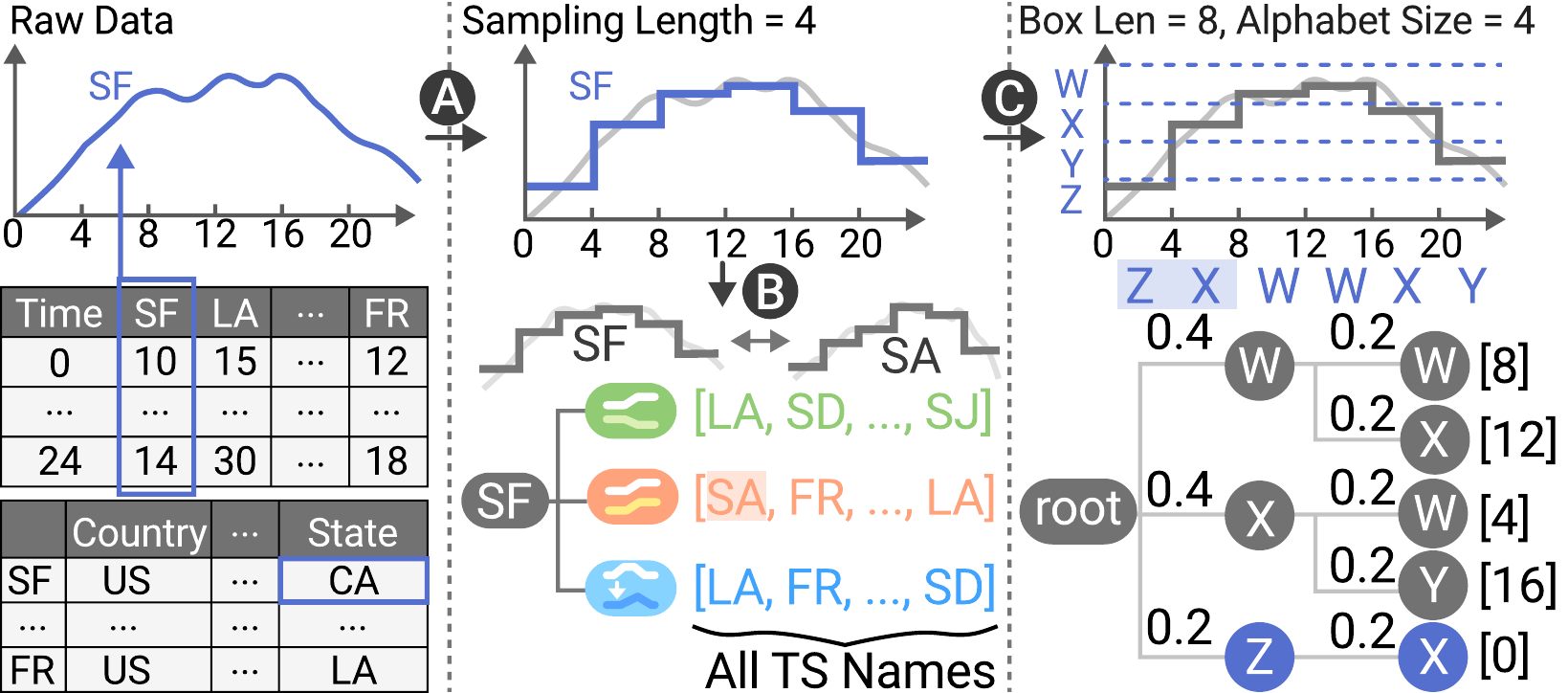}
  \caption{
     This shows a three-step data preprocessing method using an example.
     The raw data contains multiple time series (\textit{SF, LA, ..., FR}) and label descriptions (\textit{SF's State being CA}).
     (A) First, time series are compressed by taking the average of segments with Sampling Length=4 so the blue line shows compressed data.
     (B) Second, relation indexes are computed.
     We compute the relation strength between each pair of time series and record all time series names in the descending order of relation strength on the whole length.
     e.g., \textit{SA is the city with the highest similarity strength with SF}.
     (C) Third, trend indexes are computed.
     We transformed compressed data into symbolic sequences using SAX~\cite{sax} (alphabet size = 4).
     The sequence (\textit{ZXWWXY}) is built as a trie (\textit{depth = 8/4 = 2, box length = 8, sampling length = 4}).
     e.g., \textit{The starting points of ZX contain [0], and the ratio is 0.2}.
  }
  \label{fig3.1.preprocess}
\end{figure}
\subsection{Preprocessing Data}
Before querying, users should upload two CSV files - one dataset and one configuration file.
The dataset file consists of rows representing time points, with the first column indicating timestamp and each of other columns containing data for time series.
The configuration file describes labels (e.g., SF's State is CA) for each time series in each row.
RelaQ preprocesses the uploaded dataset with the following three steps (see Fig.~\ref{fig3.1.preprocess}).

First, we normalize and compress time series (Fig.~\ref{fig3.1.preprocess}A).
Users should specify \underline{Sampling Length}, the length of each segment of the time series divided for compression.
RelaQ uses PAA~\cite{paa}, which takes the average of segments to compress time series.
Second, we calculate relation indexes (Fig.~\ref{fig3.1.preprocess}B).
We enumerate pairs of compressed time series and calculate their correlation, similarity, and causality strength via methods listed in TABLE~\ref{tab1.scope} on the whole length.
All time series' names are recorded in the descending order of relation strength.
Third, we calculate trend indexes (Fig.~\ref{fig3.1.preprocess}C).
We repeat the first and second steps but using Z-normalization instead of min-max-normalization and transform data into symbolic sequences via SAX~\cite{sax} (alphabet size = 4, the size of symbols' set).
Users should specify \underline{Box Length}, the desired length of time segments the pattern should continue.
Symbolic sequences are then divided into many segments.
The division adopts a sliding window technique: the sliding step is one symbol, while the window size depends on the box and sampling length (e.g., $8/4=2$, round when indivisible).
Symbolic segments are built into a trie, where each node records possible next symbols and their occurrence ratio.
The segment starting is also recorded at the leaf node.

If the preprocessing takes longer than 2 minutes, it will be moved to the background for continued computation, allowing the user to begin querying.
When a query involves uncomputed indexes, the computation of those indexes will be prioritized.

\textbf{Justification.}
There are two types of parameters in RelaQ, pre-defined (alphabet size) and user-specified (sampling length and box length).
The selection of alphabet size considered the user's mental burden and the refinement.
We choose $\alpha=4$, trying to distinguish not only increasing/decreasing trends but also distinguish steep/gentle slopes.
As for user-specified parameters, manual clarification or automated adaptive matching would be a trade-off.
On the one hand, when the time step unit of the dataset has a clear semantic
meaning, specifying a definite length can make the query clear.
For example, if the time step unit of the dataset is ``days'', querying the daily air pollution index for a week would require a box length of ``7 days'' and a sampling length of ``1 day''.
In this case, the user-specified parameters have concrete semantics, and a slight variation to 5 days may lead to ambiguity since there is a difference between weekdays and a whole week.
On the other hand, when the semantic impact of the time step unit is not significant, especially when precision is crucial for queries, such as in milliseconds, the length of the timebox may affect the final query results.
As the overall design of RelaQ aims to minimize vague constraints in query specification and enable users to describe a query more directly, we have adopted a design that allows users to choose the length by themselves.

\subsection{Formulating Queries} 
The first step to serve easy query is enabling intuitive specification of query constraints (\textbf{R1}).
Constraints in RelaQ consist of two parts: inner-constraints (trends: \textit{the variation of a time series}, names: \textit{each name maps a specific time series}, values: \textit{the value domain of a time series}) that reveal inner features of a single time series and inter-constraints (\textit{relations}) that indicate relation features among multiple time series.

\textbf{Basic settings.}
RelaQ uses two axes settings for organizing queries: the horizontal axis encodes a timeline, and the vertical axis is divided into multiple time series tracks.
Relations are encoded using color hue and stroke width to encode the type and strength, respectively.
RelaQ supports two-mode searches to persist flexibility.
The \underline{fuzzy search} allows partial matching of constraints and slight differences, while the \underline{strict search} requires all constraints to be satisfied.
RelaQ commonly matches strictly unless users check the fuzzy option.

\begin{figure}[b]
  \centering
  \includegraphics[width=\linewidth]{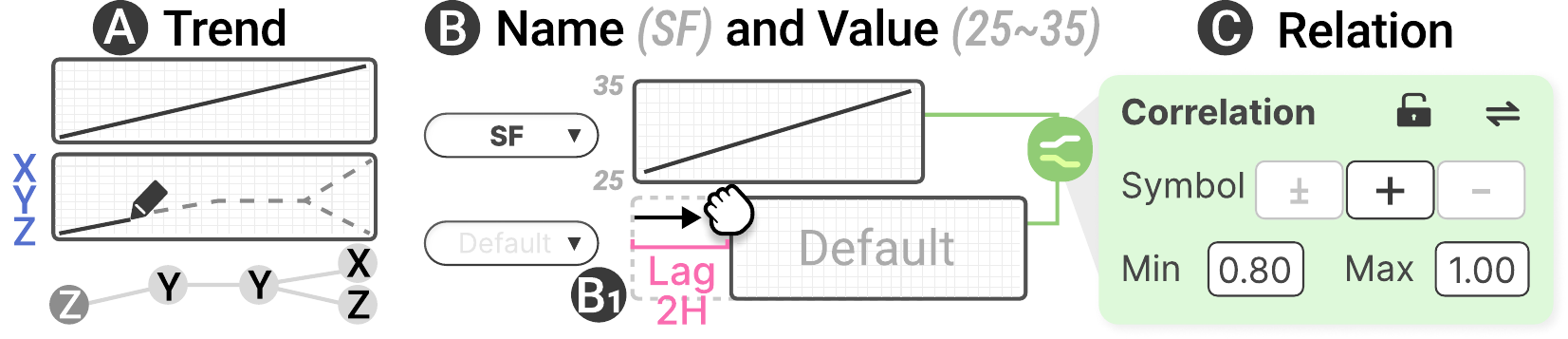}
  \caption{
  An example query illustrates inner- and inter-constraints.
  ``Which city's air quality index (AQI) will rise two hours later, following an upward trend of SF's AQI ranging in [25,35], and have a correlation strength greater than 0.8? ''
  The inner-ones include (A) rising (trend), (B) SF (name), and [25,35] (value). 
  The inter-ones include (C) ``2 hours later'' (lag) and greater than 0.8 (correlation). 
  The phrase ``which city'' implies a default timebox.
  % \update{Moreover, we need to clarify that this example is to help understand. 
  % In real processing, all time series are normalized so the range is [0,1].}
  }
  \label{fig4.query}
\end{figure}

\textbf{Inner-constraints.}
RelaQ employs \underline{timeboxes}, following the approach in~\cite{timesearcher2002}, to encode inner-constraints including trends, names, and values.
Users can sketch trends as straight line segments on meshed timeboxes (Fig.~\ref{fig4.query}A) and select names in the interface (Fig.~\ref{fig4.query}B).
Once the name is selected, the value range is determined by the minimal and maximal value of the name's corresponding time series.
The aspect ratio of sketched lines is fixed under the strict match mode, while it only indicates trends (e.g., increasing) under the fuzzy match mode.
To reduce users' mental burden, RelaQ offers two features.
First, it recommends frequent trends via trend indexes and displays them as dashed lines in the timebox, which change dynamically when sketching (Fig.~\ref{fig4.query}A).
Second, all inner-constraints can be empty and will be matched automatically.
Empty timeboxes with no inner-constraint are called default timeboxes.
A default timebox is useful when a user wants to emphasize the relation.
For example, \textit{``querying two positively correlated time fragments''} only emphasizes their relation should be the correlation but ignores inner-constraints.

\textbf{Inter-constraints.}
RelaQ facilitates the explicit specification of relations as outlined in Sec.~\ref{sec:scope}, with relation icons listed in the input panel (Fig.~\ref{fig1.teaser}A) for easy selection and addition of relation constraints.
Colored lines with relation icons are the visual representation of inter-constraints, linking timeboxes.
Since these lines display the relations between time series, we named them \underline{relalinks}.
The relation interactions cover six types summarized above and support free combinations.
Being different from the other five relations, the lag relation pertains to the time dimension, so we allow users to set lags by dragging timeboxes horizontally (Fig.~\ref{fig4.query}B$_1$) instead of clicking a certain icon.
Users can also adjust thresholds and customize options (Fig.~\ref{fig4.query}C) to allow fuzzy matching and reverse matching (e.g., minimum similarity).

\textbf{Justification.}
We mainly justify the design alternatives of constraint specification interactions. 
As discussed in Sec.~\ref{sec:related_work}, Mannino et al.~\cite{qetch} identified that there were three types of sketch approaches to query time series: a) overlay sketches, b) annotated sketches, and c) restricted sketches.
Overlay sketches rely on a specific reference time series, so the name must be specified, limiting the flexibility.
Annotated methods can lead to visual clutter when multiple time series with labels are not placed clearly and organized.
In contrast, restricted sketches~\cite{Keogh1997straight} focus on capturing the essential features.
Restricted sketches can make the sketch more visually appealing and easier to understand for users at a glance.
We take clarity and simplicity as key considerations when designing the specification approach, so we finally adopt restricted sketches.
\subsection{Processing Queries}
\begin{figure*}[htbp]
  \centering
  \includegraphics[width=\linewidth]{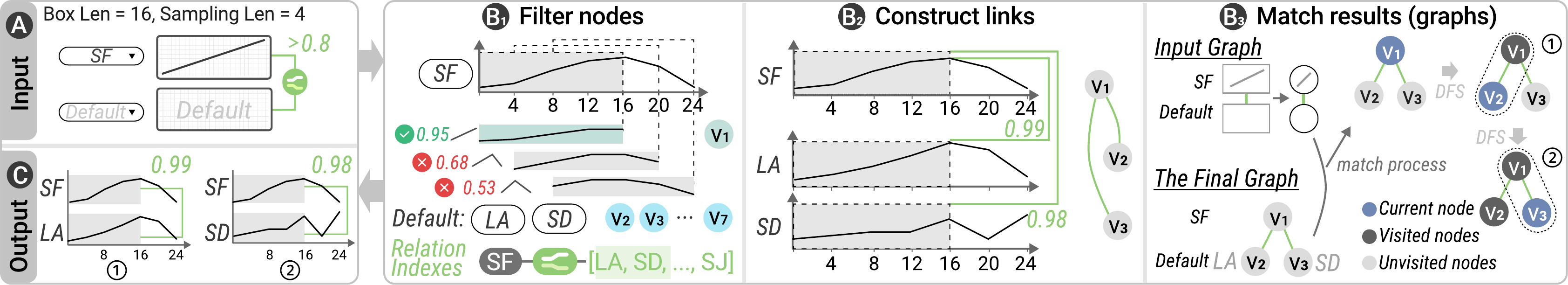}
  \caption{
     This figure displays a simplified example query: \textit{``Which city has a temperature that rises together with San Francisco and has a correlation strength above 0.8''}
     (A) The user input includes two timeboxes, [name (SF), trend (rising)] and [default (which city)], as well as relations, [correlation (above 0.8)] between SF and the default timebox. 
     (B$_1$) RelaQ filters time fragments based on trends and records them as nodes. $v_1$ is the only node that meets the rising trend with a degree of 0.95 in SF, while $v_{2-7}$ are all valid for the default timebox. 
     (B$_2$) For all pairs (SF-default) of nodes, relations are checked. The correlation strength between $v_1$ (SF) and $v_2$ (default) is more than 0.8, so we construct a link. The same is between $v_1$ and $v_3$. 
     (B$_3$) RelaQ uses a depth-first search on the final graph and obtains two results.
     (C) Two results reveal that SF will affect LA and SD with a strength of 0.99 and 0.98, respectively.
  }
  \label{fig5.algorithm}
\end{figure*}
     % a) The input query is abstracted into a node-link graph $G=(V,E)$, where $V = \{v_1,v_2\}$ and $E = \{e_1=(\pm0.8,2)\}$.
     % b) The query model loads a dataset that is composed by three time series (A, B, C).
     % The length of timebox is 6 and the length of series is 9, so there are 3 nodes for each series.
     % The final result graph $G'=(V',E')$, where $V'=\{v_{11}', v_{21}', v_{22}'\}$ and $E' = \{e_{11}', e_{12}'\}$.
     % c) The output contains two answers. The score of first alternative is $0.87 + |0.99| + 1.00 = 2.86$.
Besides the aforementioned specification approaches, an effective model that allows matching heterogeneous constraints flexibly is also necessary to support easy query (\textbf{R1}).
The query process is end-to-end: Every time a user edits a query, RelaQ searches for all matched results via a three-step matching algorithm.
An example input and output are displayed in Fig.~\ref{fig5.algorithm}AC, while Fig.~\ref{fig5.algorithm}B$_{1-3}$ reveal the matching process.

Since dynamic relations that change over time are often modeled as graph problems, we are inspired to leverage the node-link graph as a proxy to model this matching problem.
For each input graph, nodes are timeboxes, and links are relalinks.
To distinguish, in the output graph, nodes are fragments, and links are relations.
Thus, this problem is to match the input graph on the constructed dataset graph.
 
The construction of the dataset graph and matching are integrated and progressive.
The method is three-step (Fig.~\ref{fig5.algorithm}B): (1) filter valid fragments as nodes, (2) construct links based on relations, and (3) match final results.
To simplify the narration, we illustrate the method with an example of a single non-dynamic relation. 
For more detailed explanations of dynamic-relation queries, please refer to the supplementary material.

First, RelaQ filters fragments based on timeboxes $x_i$.
For each timebox, we obtain the compressed time series according to its \textit{name} and use a sliding window technique to divide it into several fragments.
The window size is the box length, while the sliding step is the sampling length.
Then, we measure the \textit{trend} matching degree based on ED~\cite{ed}.
If the degree exceeds threshold $s=0.7$, we record the fragment as a valid node in the set $X_i$. 
For example, regarding timebox $x_1$ belonging to SF, only $v_1\in X_1$ matches the rising trend with the degree $0.95$ in Fig.~\ref{fig5.algorithm}B$_1$.
When the trend constraint is not declared, the matching degree will be set as $1$.
For the default timebox with no name, RelaQ enumerates the first 20 time series in the relation indexes.
For instance, in Fig.~\ref{fig5.algorithm}B$_1$, LA and SD are the first two cities with the highest correlation to SF.

Second, RelaQ builds links based on relalinks $y_k$, which links $x_i$ and another timebox $x_j$.
We obtain $X_i$ and $X_j$ valid nodes computed in the first step.
For each relation $y_k$, we validate each pair of compliant fragments (nodes) in $X_i \times X_j$, checking lag constraints and computing relation strength on the fragment length via discussed formulas in TABLE~\ref{tab1.scope}.
If the relation strength of a pair is over user-given threshold of $y_k$, we reserve it as a valid link in the set $Y_k$.
For instance, the relation strength of link $e_1 \in Y_1$ between $v_1 \in X_1$ and $v_2 \in X_2$ is $0.99$ in Fig.~\ref{fig5.algorithm}B$_2$.
The final dataset graph is $G = (X,Y)$.

Third, RelaQ uses depth-first search~\cite{tarjan1972depth} with memoization pruning techniques to form results.
The search process is to match the input graph in the built dataset graph.
The search order is based on the temporal order of timeboxes.
RelaQ starts the search from the earliest timebox in a query.
Each result is a connected subgraph $g = (v, e)$ in $G$.
In Fig.~\ref{fig5.algorithm}B$_3$, the search starts from $v_1$, then transfers to $v_2$, forming the first result, and ends on $v_3$, forming the second result.

Besides, the matching score $r(g)$ is defined as Eq.~\ref{eq:rg}, here $d(v)$ is the trend matching degree of node $v$, and $l(e)$ is the relation strength of link $e$.
In Fig.~\ref{fig5.algorithm}C, the matching score of the first result is $0.95 + 1 + |0.99| = 2.94$.
\begin{equation}
    r(g) = \textstyle\sum_{v\in X} d(v) + \textstyle\sum_{e \in Y}  |l(e)|
\label{eq:rg}
\end{equation}

\subsection{Interpreting Results}
Intuitive visualizations of queried results are critical to helping users understand retrieved time series and underlying patterns.
To satisfy \textbf{R2}, RelaQ should (1) visualize queried time series with relations and (2) enable comparing different results flexibly.
RelaQ adopts a matrix view that compactly organizes results in a matrix for comparison and a time view that depicts results along the timeline in a temporal context.
\begin{figure}[ht]
  \centering
  \includegraphics[width=\linewidth]{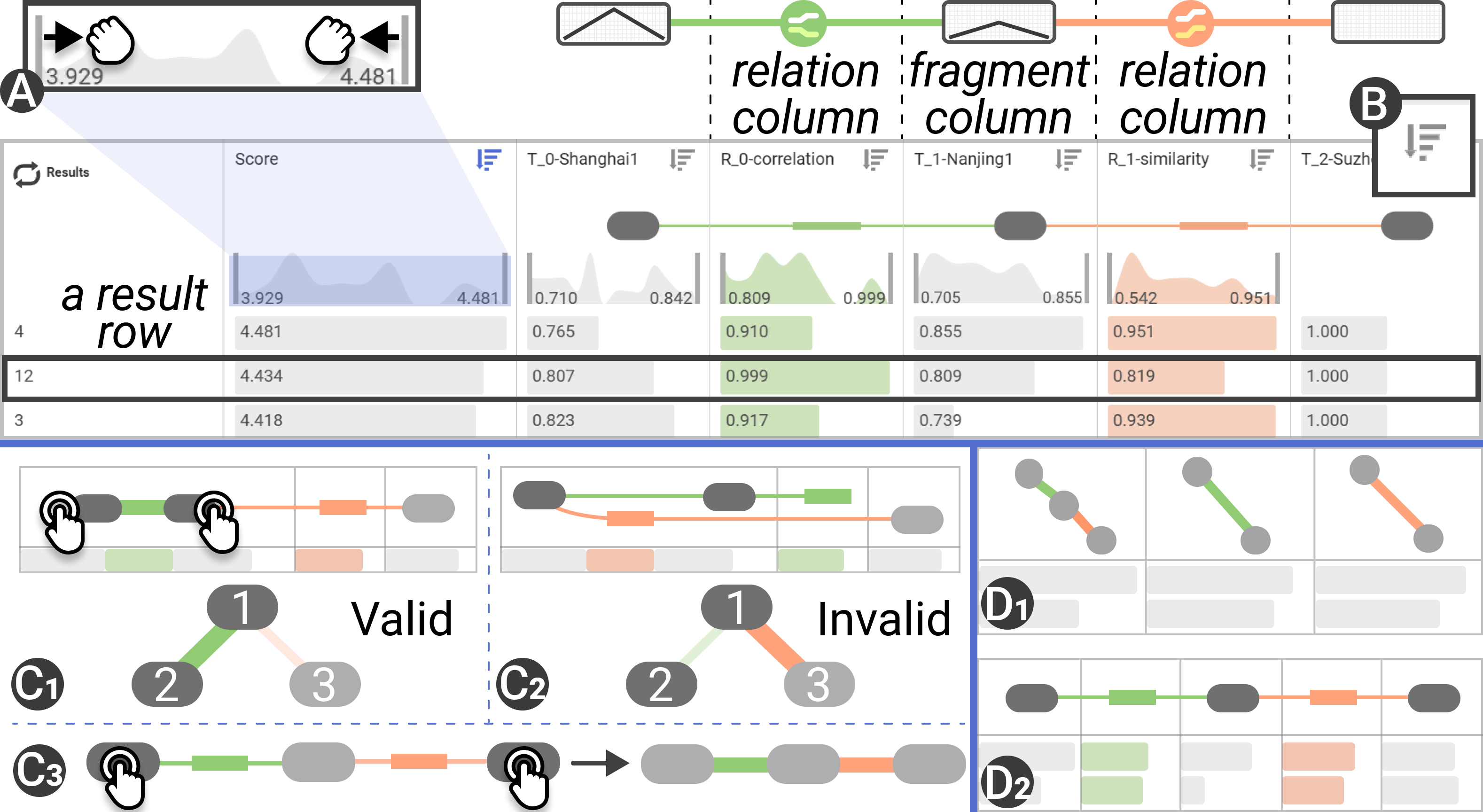}
  \caption{
  The visual design of the matrix view.
  Each row of the matrix is a result and each column matches a relation or a fragment.
  There are three interactions: (A) filter the range of distribution, (B) rank the results by a specific column, and (C$_{1-3}$) combine some columns.
  (C$_1$) shows a valid combination, a green relation links fragment1 and fragment2, while (C$_2$) is invalid, where the orange relation links these two fragments incorrectly.
  (C$_3$) reveals the auto-combination feature of RelaQ.
  (D$_1$) is a design alternative based on juxtaposed graphs
   and (D$_2$) is the dismantled graph we adopt.
   }
  \label{fig6.1.matrix}
\end{figure}

\textbf{The matrix view (Fig.~\ref{fig6.1.matrix})} is inspired by LineUp~\cite{Lineup}, a widely used technique for comparing multidimensional items.
In the matrix view, each row is a result and each column is a fragment (the result of a timebox) or a relation (the result of a relalink).
Since fragments and relations together display a pattern, we connect the columns of fragments and relations with links to keep consistency with the input query.
The column header displays the distribution of strengths for relations or trend-matching rates for fragments.
Users can adjust the range of distribution to filter results by moving the slider (Fig.~\ref{fig6.1.matrix}A).
Each column has a sorting icon that (Fig.~\ref{fig6.1.matrix}B), when clicked, will arrange the results in descending order based on the matching score of that column.
Clicking the icon again will switch to ascending order.
To further assist users in comparing patterns that have multiple relations and fragments, we allow them to combine multiple columns.
RelaQ enables users to combine columns by simply clicking on them.
For instance, see Fig.~\ref{fig6.1.matrix}C$_1$, users can click the first three columns (two fragment columns and a correlation column) to view a sub-pattern, such as the correlated portion of a query.
However, some combinations may not be valid (Fig.~\ref{fig6.1.matrix}C$_2$), that is, when relations and fragments are linked incorrectly, patterns may be meaningless.
Therefore, RelaQ employs an auto-combination mechanism to ensure meaningful sub-patterns are grouped.
When users combine any two fragments, RelaQ will find all relations and fragments between these two fragments and combine them all.
As shown in Fig.~\ref{fig6.1.matrix}C$_3$, when the user clicks on the two end fragments, all the fragments and relations that connect them in between will be combined.
\begin{figure}[b]
  \centering
  \includegraphics[width=\linewidth]{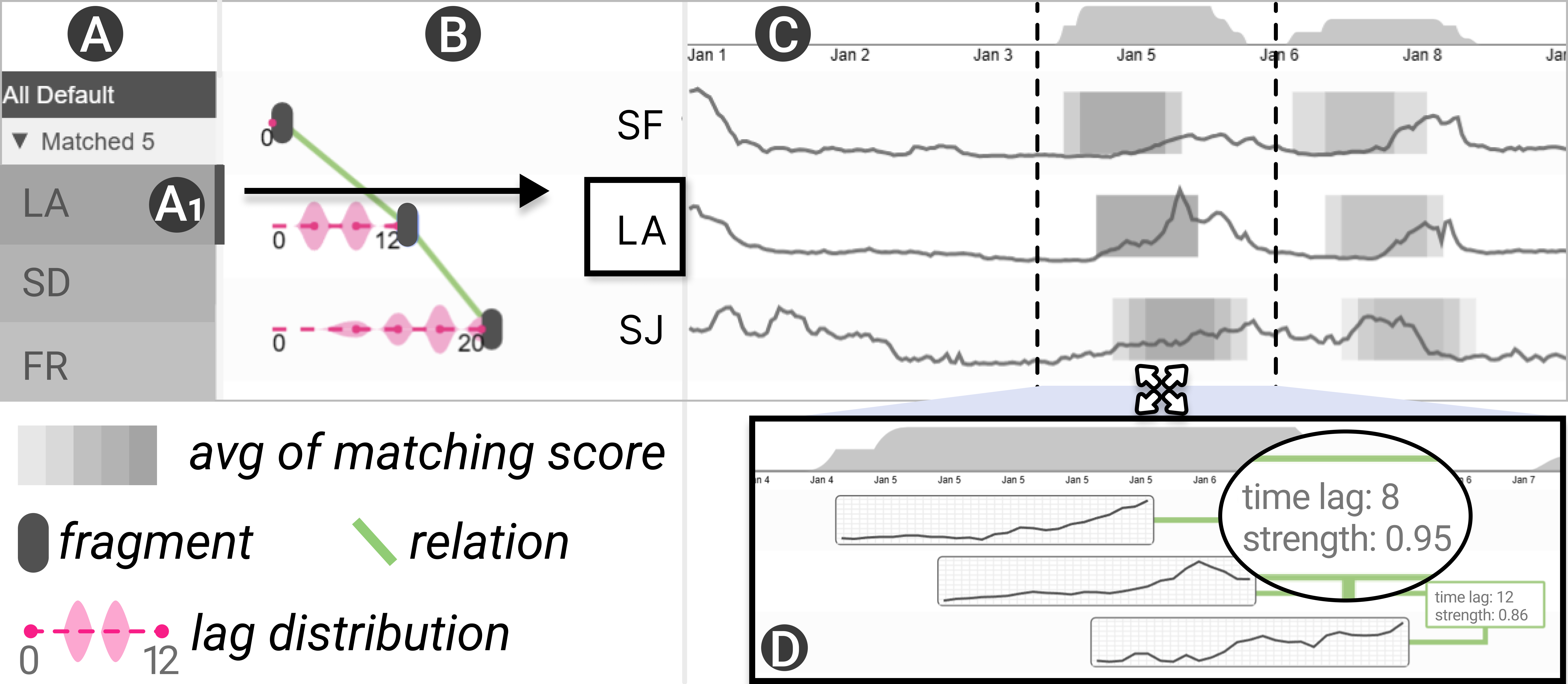}
  \caption{
  The visual design of the time view.
  (A) is the list of alternative time series matching the default timebox in the decreasing order of its average matching score.
  (B) is an overview of all results.
  (C) is the distribution of results and raw time series.
  (D) shows the detail level of a result.}
  \label{fig6.2.time}
\end{figure}

\textbf{The time view (Fig.~\ref{fig6.2.time})} consists of three parts arranged from left to right.
First, as the default timebox has not been specified with a name, RelaQ presents a list of time series (Fig.~\ref{fig6.2.time}A) that can be used to complete the default timebox.
For example, SF-default can be completed with either SF-LA or SF-SD, and both LA and SD will appear in the list.
The opacity of each item in the list encodes the average matching score (divided by the highest one) when completing the default timebox with that particular time series.
If a time series is selected to complete the default timebox, it will be highlighted like Fig.~\ref{fig6.2.time}A$_1$.
Second, the structure overview of all results is visualized as a node-link graph (Fig.~\ref{fig6.2.time}B).
The thickness of lines encodes the average strength of such relations.
We also visualize the time lag distribution between two fragments.
Third, the time series involved in the queried results are visualized with line charts in different rows (Fig.~\ref{fig6.2.time}C).
An area chart on the top shows the temporal occurrence of the results.
Matched time fragments are highlighted in the line chart to make the matched pattern clear.
Users can zoom in on each result to obtain details (Fig.~\ref{fig6.2.time}D).
RelaQ adopts the same timebox visualization as the input panel for visual consistency and easy comparison.

\textbf{Justification.}
There are many approaches supporting comparison of node-link graphs according to Beck's survey~\cite{graphsurvey}.
We evaluate two types of visualization techniques, namely juxtaposed and dismantled graphs.
In juxtaposed graphs (Fig.~\ref{fig6.1.matrix}D$_1$), each sub-pattern is presented as a column, while in dismantled graphs (Fig.~\ref{fig6.1.matrix}D$_2$), nodes and relations are split into separate columns.
Many visual analytics applications employ juxtaposition techniques (e.g., \cite{largenetvis,snap,compass}), which is useful for comparing subgraphs.
However, juxtaposition techniques suffer from scalable problems and do not allow adjusting the weight of nodes and links.
Thus, we adopt dismantled graphs.

\subsection{Providing Recommendations}
The guidance panel instructs users when they want to perform a breadth-oriented query (\textbf{R3}).
It provides alternative timeboxes that can be added to an existing query, just like the ``related searches'' module in the bottom of a search engine results page.
The guidance panel supports displaying relevant timeboxes that can be added to the existing query and comparing alternative timeboxes.

\begin{figure}[hb]
  \centering
  \includegraphics[width=\linewidth]{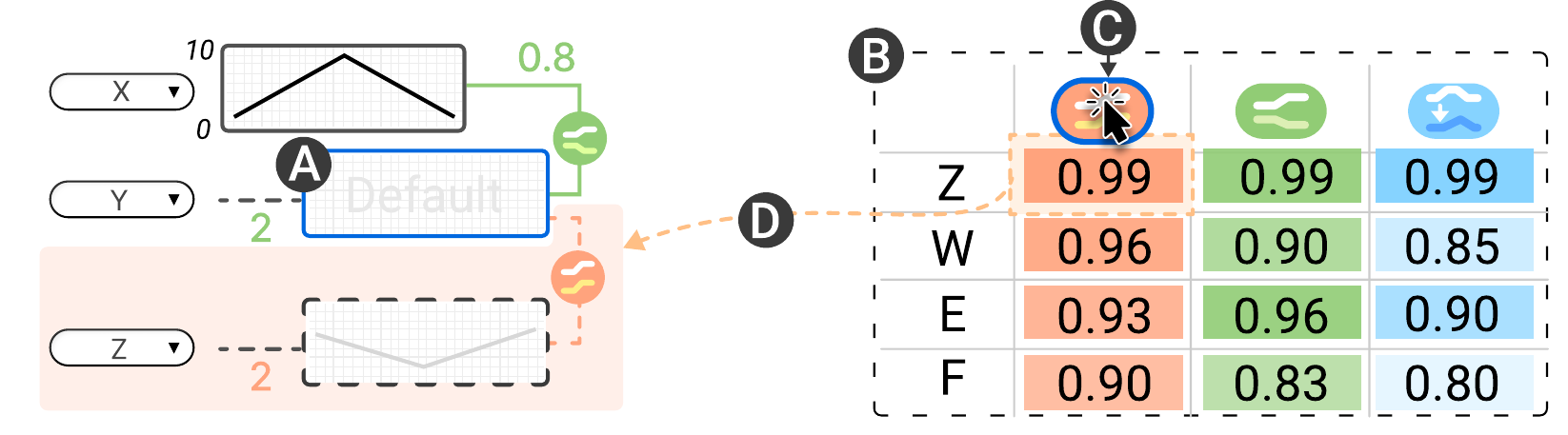}
  \caption{
    The guidance panel's design.
     (A) The focus timebox. 
     (B) The recommended timeboxes that can be added to the query.
     (C) The relation icon that can be clicked to sort recommendations.
     (D) The preview of a new query after hovering.
  }
  \label{fig7.guide}
\end{figure}

\textbf{Visual design.} Presenting all recommended timeboxes simultaneously may not be feasible due to the large number of potential recommendations. 
To address this issue, we propose a solution that involves the provision of recommendations on demand. First, we require users to specify their ``focus timebox'' (Fig.~\ref{fig7.guide}A) by hovering.
We exclusively search for time series that are relevant to the ``focus timebox''.
Second, we employ a matrix-based design (Fig.~\ref{fig7.guide}B) to display all recommendations.
Each cell in the matrix represents an alternative timebox that can be added to the existing query.
Cells within the same row pertain to a single time series, while cells within the same column denote the same type of relation.
Users can click the icon in the column header to sort timeboxes (Fig.~\ref{fig7.guide}C).
We utilize a confidence-based approach to determine the order of cells.
Specifically, the confidence is calculated based on the percentage of the relation within the total recommended relations.
This approach ensures that recommendations are sorted based on their relevance and usefulness to the user.
The opacity of each cell encodes the confidence, while the text means the strength.
We allow users to hover over a cell and check its preview (Fig.~\ref{fig7.guide}D).

\textbf{The recommendation algorithm.}
This explicitly features three relation types: similarity, correlation, and causality, which we selected based on their complexity and dynamic nature.
Besides, the relation lag is also recommended implicitly with these three relations.
We adopt a statistical recommendation algorithm.
A group of time series related to the ``focus timebox'' were selected from pre-calculated relation indexes.
RelaQ then evaluates each time series by enumerating lags and summarizes the average strength and confidence.
Finally, the top 20 time series with the highest confidence are displayed.
Relations, including meta and arithmetic, were not included in the analysis due to their complex possible scenarios and large search space, which can significantly impact the algorithm's time performance. Moreover, compared to other relation types, they are not commonly queried.
Thus, RelaQ does not support recommendations for meta and arithmetic relations.

\section{Case Study}
This session is divided into three stages: \textbf{[20 min]} a 15-minute tutorial and 5 minutes free exploration, \textbf{[50 min]} Free relation-driven retrieve of time series based on real-world dataset (Q-Queries, R-Results), following the think-aloud protocol\cite{fonteyn1993description}, \textbf{[10 min]} a final one-to-one expert interview.
The aim of the case study is to (1) evaluate the effectiveness and usability of RelaQ and (2) collect feedback from experts.

\subsection{Case 1: Analyze the relation between brain regions}
\label{sec:case1}
We invited Expert A (female, EA), who has four years of experience in analyzing EEG data.
EA aimed at figuring out how the correlation between two brain regions changed after drinking when humans were exposed to two matched stimuli.

\textbf{Dataset.}
This case study is based on an \href{https://archive.ics.uci.edu/ml/datasets/eeg+database}{EEG Database} data set, which includes the control group data and the alcoholic group data.
Each group tests eight subjects under two matched stimuli and collects signals under 256 Hz in a second.
To reduce noises, we average multiple results from the same subject under the same condition.
The final data set has 122 * 256 = 31,232 samples and a size of 275KB.
Fp1, Fp2, Fpz, and AF7 are \href{https://www.bing.com/images/search?view=detailV2&ccid=alQn1OYy&id=CF8DCBF27AEEB87C9BD68DCBE08ACCC5229F528A&thid=OIP.alQn1OYyleV3YbwdjPv4CQHaHG&mediaurl=https%3A%2F%2Fwww.researchgate.net%2Fprofile%2FJose_Del_R_Millan%2Fpublication%2F330340445%2Ffigure%2Fdownload%2Ffig1%2FAS%3A714229665517574%401547297030604%2F64-EEG-Electrodes-layout-extended-10-20-international.ppm&cdnurl=https%3A%2F%2Fth.bing.com%2Fth%2Fid%2FR.6a5427d4e63295e57761bc1d8cfbf809%3Frik%3DilKfIsXMiuDLjQ%26pid%3DImgRaw%26r%3D0&exph=815&expw=850&q=EEG+Electrode+Map&simid=608028345799686813&form=IRPRST&ck=F0684EFB3481079050E1240CFF2B19FB&selectedindex=1&ajaxhist=0&ajaxserp=0&vt=0&sim=11}{electrodes}. (c: common, a: alcoholic)

\begin{figure}[t]
  \includegraphics[width=\linewidth]{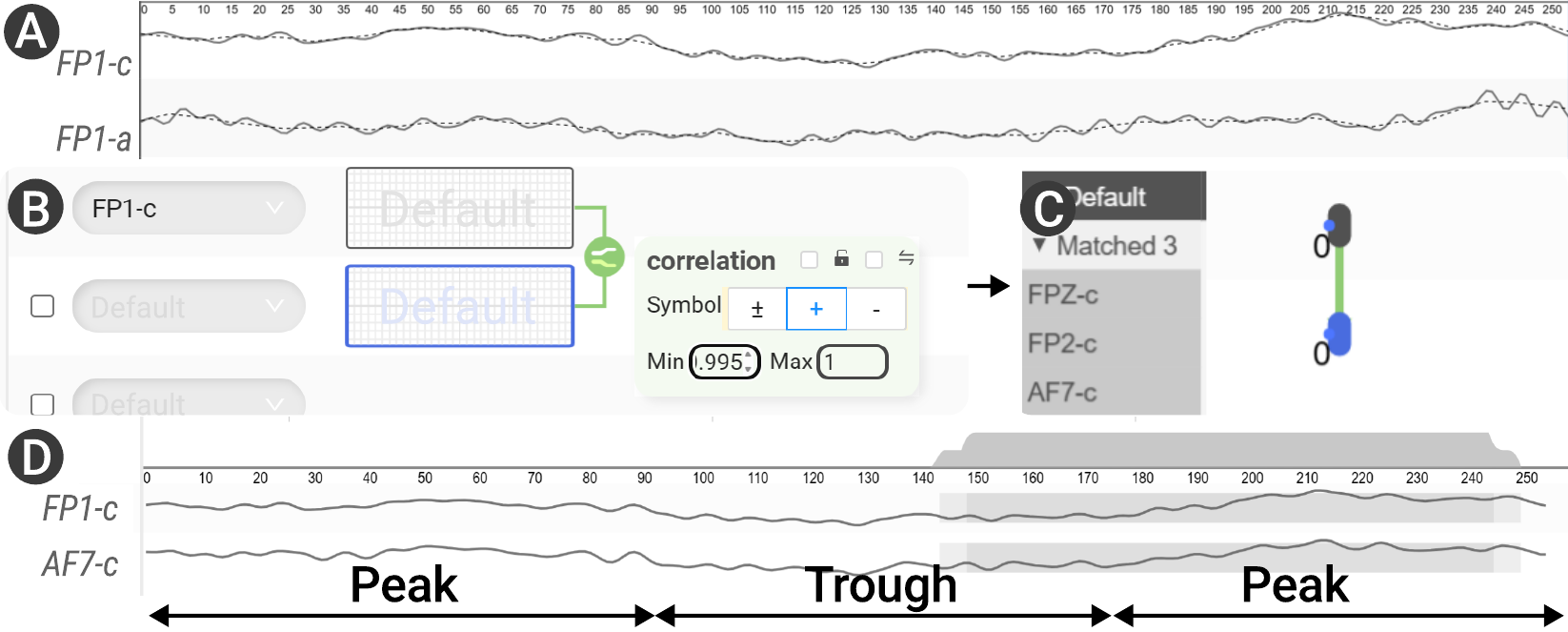}
  \caption{This displays Q1 and R1 for the first case. (A) The time series of Fp1-c and -a. (B) Query time series correlated to Fp1-c. (C) Results include FPZ-c, FP2-c, and AF7-c. (D) Results mainly distribute in the second peak. }
  % \Description{}
  \label{fig9.case}
\end{figure}

\textbf{Q1. Query which electrodes correlated with Fp1 before drinking.} % overview 1st query
EA first loaded the data and adjusted Sampling Length (5) and Box Length (100).
The difference between time series of Fp1-c and Fp1-a (Fig.~\ref{fig9.case}A) attracted EA as it indicated that alcohol affected Fp1.
Thus, EA began to query the electrodes correlated to Fp1 before drinking.
She created two timeboxes, identified one as Fp1-c, and kept the other as default.
Then, she selected correlation and clicked two timeboxes to link them.
EA set the threshold as [0.995,1] according to domain knowledge, as shown in Fig.~\ref{fig9.case}B.

\textbf{R1. Explore how electrodes correlated with Fp1 before drinking.} % insight of 1st query
EA found there were three time series in the results (Fig.~\ref{fig9.case}C), including Fp2-c, Fpz-c, and AF7-c.
She said that \textit{it was reasonable because Fp2 and Fpz were in the same region with Fp1, while AF7 was close to Fp.}
The result was in line with her expectations.
Furthermore, when EA inspected details, she found that results with a high matching score of AF7 were most in the right peak segment (Fig.~\ref{fig9.case}D).
This phenomenon inspired her that the effect might vary in the peaks and troughs.
Thus, EA filtered the results and found that results with high correlation were mainly distributed in the peaks for all three pairs of electrodes.
She said it indicated that these three pairs of electrodes show synergy to this stimulus and the peak might represent the response.

\textbf{Q2. Query the pattern of correlation changes after drinking.} 
Next, EA continued to query how the correlation strength between Fp1 and other electrodes decreased after drinking.
She created a timebox labeled Fp1-a, linked it to a new default one, and set the correlation relation constraint.
Then, EA added the meta relation constraint of two default timeboxes.
This requires that two timeboxes (Fig.~\ref{fig1.teaser}D$_1$ and D$_2$) be from the same electrode (Fig.~\ref{fig1.teaser}D$_3$).
D$_1$ has a high correlation with Fp1-c, while D$_2$ has a low correlation with Fp1-a (Fig.~\ref{fig1.teaser}D$_4$ revert matching finds minimum value). 
Thus, the query selects the electrode whose correlation with Fp1 decreases the most.
The query is shown in Fig.~\ref{fig1.teaser}D.
After a few seconds, EA found that Fp2, Fpz, and AF7 were still in the results, and Fp2 was affected most (Fig.~\ref{fig1.teaser}E).
EA scrolled to inspect several results. 
EA said, \textit{``The seriously decreased correlation strength between Fp1 and Fp2 may indicate some some asymmetrical impacts.''}

\textbf{R2. Explore differences between the peak and the trough.}
Inspired by the observation in the second step, EA further analyzed whether the difference between the peak and the trough still existed.
Thus, she checked the results' distribution in the Fp1-Fp2 pair.
EA filtered results with higher matching scores and found the time range of the peak was highlighted obviously (Fig.~\ref{fig1.teaser}F).
The Fp1-Fpz and Fp1-AF7 pair had the same visual pattern, which EA thought was interesting.
It revealed that the peak region was affected the most.
To investigate details, EA observed time series in the analysis panel.
She found that the peak appeared delayed in the alcoholic group (Fig.~\ref{fig1.teaser}G).
Till now, EA could summarize some preliminary results.
She added, \textit{``The correlation between Fp1-Fp2 decreases most and the peak delays reveal that alcohol may affect balance ability and reaction speed, respectively.''}

\subsection{Case 2: Explore the relation of air pollution between cities in the North China Plain}
Analyzing air quality time series is a popular and important topic~\cite{compass}, so we invite Expert B (male, EB), who has five years of experience in analyzing urban air data.
Many efforts have been devoted to protecting the environment in the North China Plain, which is a densely populated area and constantly affected by air pollution.
Since Beijing was one of the most important monitoring sites located there, EB explored the air pollution patterns involving Beijing.

\textbf{Dataset.} The dataset collects the PM2.5 concentration from 448 air quality monitoring sites in China.
Each site records hourly between Jan 1$^{st}$ and Dec 20$^{th}$, 2018.
This dataset has $448*8,472 = 3,795,456$ records and a size of 11.4MB. 
\begin{figure*}[htb]
  \includegraphics[width=\linewidth]{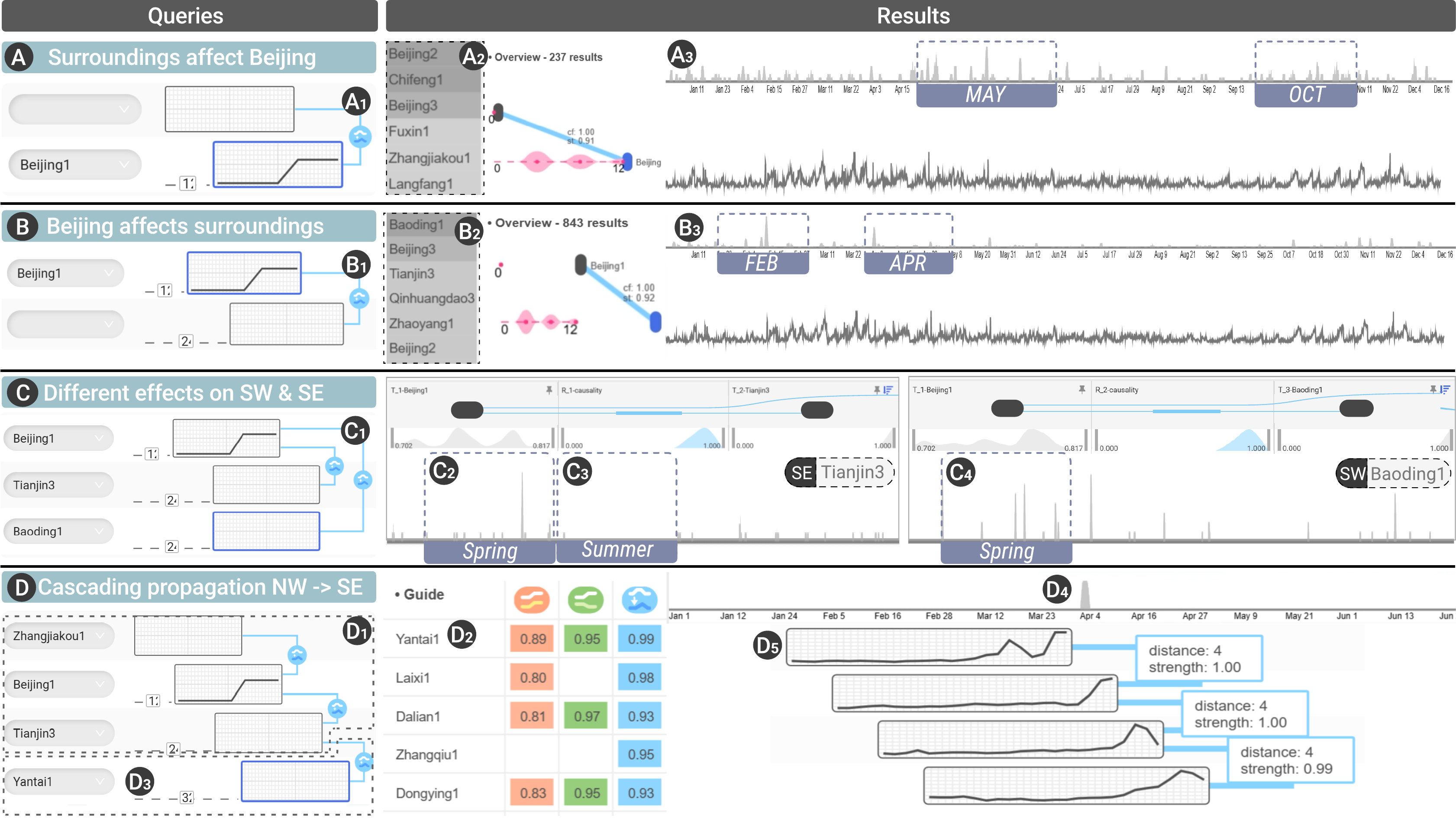}
  \caption{4 queries and results in Case 2.
(A) How does Beijing's surroundings affect it?
(B) How does Beijing affect its surroundings?
(C) What are the differences in pollution transmission from Beijing to various directions?
(D) What is the pattern of cascading propagation from Zhangjiakou to Beijing?}
  % \Description{}
  \label{fig10.case2}
\end{figure*}

\textbf{Q1R1. Query how surroundings affected Beijing.} % 1st query
The Sampling Length (4H) and Box Length (24H) were adjusted at the beginning.
First, EB was curious about how air pollution spread from surrounding cities to Beijing.
EB added a timebox A (Default) and a timebox B (Beijing1), and clicked the causality icon to connect them.
He then dragged B nearly to the half of A because the delay in propagation might be up to 12 hours depending on the wind speed.
Since EB was interested in when the air pollution index is rising, he sketched a trend in timebox B referencing the trend guidance.
The query is shown in Fig.~\ref{fig10.case2}A$_1$.
After a few seconds, 237 results in total have been displayed on the screen. 
EB found that Beijing was affected throughout the year and the peak occurred in May and October (Fig.~\ref{fig10.case2}A$_3$).
He explored which cities affected Beijing most.
The order of alternatives (Fig.~\ref{fig10.case2}A$_2$) revealed that Beijing has been affected by air pollution from the north (Zhangjiakou) more seriously than the south (only Langfang).

\textbf{Q2R2. Query how Beijing affected surroundings.} % 2st query
EB then wondered how Beijing's air pollution had spread, so he adjusted the query (Fig.~\ref{fig10.case2}B$_1$).
There were 843 results, which was nearly four times more than the last one.
Peaks occurred in February and April (Fig.~\ref{fig10.case2}B$_3$).
EB explained that with some effective environmental governance measures like the use of new energy, the current source of air pollution has mainly become urban traffic pollution.
Through the alternative list (Fig.~\ref{fig1.teaser}B$_2$), EB further found that the air pollution transmitted from Beijing mainly to the south (Tianjin, Baoding) and the east (Qinghuangdao) but hardly to the west.
EB explained that the western mountains blocked pollution from spreading from Beijing to the northwest regions with higher elevations.

\textbf{Q3R3. Compare the spread of air pollution in different directions.} % 2nd comprehension
Inspired by the observations in Q2, EB was interested in the differences between patterns of air pollution propagation from Beijing to different directions.
He configured Timebox A as Tianjin (southeast) and Timebox B as Baoding (southwest) (Fig.~\ref{fig10.case2}C$_1$).
To compare these two propagation patterns, EB grouped two subgraphs involving Tianjin and Baoding, respectively.
He found that the results were mainly distributed in spring for both cities (Fig.~\ref{fig10.case2}C$_{2,4}$).
Moreover, Tianjin was hardly affected in summer (Fig.~\ref{fig10.case2}C$_3$) but Baoding was affected nearly throughout the whole year.
EB said that the results were in line with climatic characteristics, \textit{``Because of monsoon factors, the wind blows from Tianjin to Beijing in summer, so there is less pollution coming to Tianjin.''} 

\textbf{Q4R4. Query cascading propagation involving Beijing.}
EB investigated the chain of air pollution propagation from northwest to Beijing and its effects.
He removed Baoding and added Zhangjiakou (Fig.~\ref{fig10.case2}D$_1$).
EB clicked the guidance icon to find cascading effects.
After the recommended queries were shown, EB triggered the sorting in the guidance panel and found Yantai had the highest average strength with the highest confidence (Fig.~\ref{fig10.case2}D$_2$).
Thus, EB added Yantai to the existing query (Fig.~\ref{fig10.case2}D$_3$).
In the results, the distribution showed that the cascading air pollution patterns mainly occurred in April (Fig.~\ref{fig10.case2}D$_4$).
EB inspected several results with the highest matching score (Fig.~\ref{fig10.case2}D$_5$) and found that they aligned with his expectations.
These findings suggest the possibility of long-range air pollution transmission from Zhangjiakou to Yantai in spring, emphasizing the need for preventative measures.

\subsection{Expert Interview}
After the experts finished analyzing the real-world dataset, we followed a semi-structured questionnaire to collect their feedback on effectiveness, usability, and improvements.

\textbf{Effectiveness and visual designs.}
Both EA and EB spoke highly of RelaQ and confirmed its effectiveness and intuitiveness.
EA commented it was intuitive to formulate her queries through RelaQ. 
\textit{``Functions including relation interactions, sketches, and default constraints support complex queries, which can cover massive analysis scenarios that previously had to be done by writing scripts,''} she said.
EB noted RelaQ as \textit{``useful''} since existing important patterns, such as cascading propagation or ego-network effects, can both be queried easily.
Moreover, EA and EB praised the matrix view in filtering, grouping, and comparing desired patterns, saying it was \textit{``flexible''} and \textit{``powerful''}.
In addition, EB commented the guidance was helpful, saying \textit{``It helps me quickly narrow down my choices from a huge number of sites.''}

\textbf{Usability and improvements.}
Though EA and EB agreed that RelaQ had good usability, noting that interactions and designs were \textit{``fluency''} and \textit{``expressive''}, they also gave many suggestions on improving RelaQ.
EA recommended adding distribution visualization of parameters in the query panel to guide users in setting initial queries.
EB suggested that we should highlight how the trend matches and support adding time series in the result view on demand.
We have optimized RelaQ based on their suggestions.

\section{User Study}
We conducted a task-based user study to evaluate the usability of RelaQ and collect usability feedback. 

\subsection{Experiment Settings}

\textbf{Participants and Data. }
We recruited twelve participants (S1-S12, six males and six females) from different departments, including Computer Science (6), Traffic Engineering (1), Electronic Engineering (1), Mathematics (1), Industrial Design (1), Media (1) and Sports Science (1).
All subjects were familiar with time series data, with an average self-reported score of 3.5 on a 5-point Likert Scale.
Among them, 7 subjects had experience using tools to query or explore time series data, including Python, Microsoft Excel, MATLAB, and Unity.
None of the subjects have been involved in the development of RelaQ.
In this study, we use a dataset (142 time series * the length 5,000) collected from a real factory.

\textbf{Procedure and Tasks. }
The experiment lasts about 50 minutes and includes the following steps.
First, subjects were presented with a 15-minute tutorial on the background of time series query and the visual encodings and interactions of RelaQ.
Then, they freely used the system for 5 minutes.
Afterward, subjects were required to complete nine tasks.
All tasks were designed to evaluate RelaQ's different functions, including specifying queries (\textbf{R1}), understanding results (\textbf{R2}), and exploring possible relations (\textbf{R3}).
We ensured that all interactions and visual encodings were covered during the process.
We divided these tasks into the following two stages according to the type of queries summarized in Sec.~\ref{sec:requirements}.

\begin{compactitem}
    \item \textbf{Stage 1. Query target relations (\textbf{R1}, \textbf{R2}).  }
    Subjects are required to specify queries (\textbf{R1}) and answer questions (\textbf{R2}).
    We asked subjects to design three queries and answer questions (T1-3) with different levels of difficulty.
    \item \textbf{Stage 2. Explore possible relations (\textbf{R2}, \textbf{R3}). }
    Subjects are required to explore possible relations (\textbf{R3}) and answer questions (\textbf{R2}).
    We asked subjects to explore based on recommended series (T4, T7) and trends (T5-6) and understand results (T8-9).
\end{compactitem}

Subjects are asked to follow the think-aloud protocol~\cite{fonteyn1993description}.
Their completion time, behavior, and answers were all recorded.
After completing tasks, subjects were required to fill in a post-test questionnaire based on System Usability Scale (SUS)\cite{SUS} with a 5-point Likert scale. 
Finally, we conducted a semi-structured interview about their experience of using RelaQ to collect feedback.

\subsection{Quantitative Results}
To faithfully reflect how users completed tasks using RelaQ, we counted the completion time and the pass rate.
We considered a subject as passed when he/she fully understood the task, performed reasonable operations, and answered correctly. 

\begin{figure}[t]
  \centering
  \includegraphics[width=\linewidth]{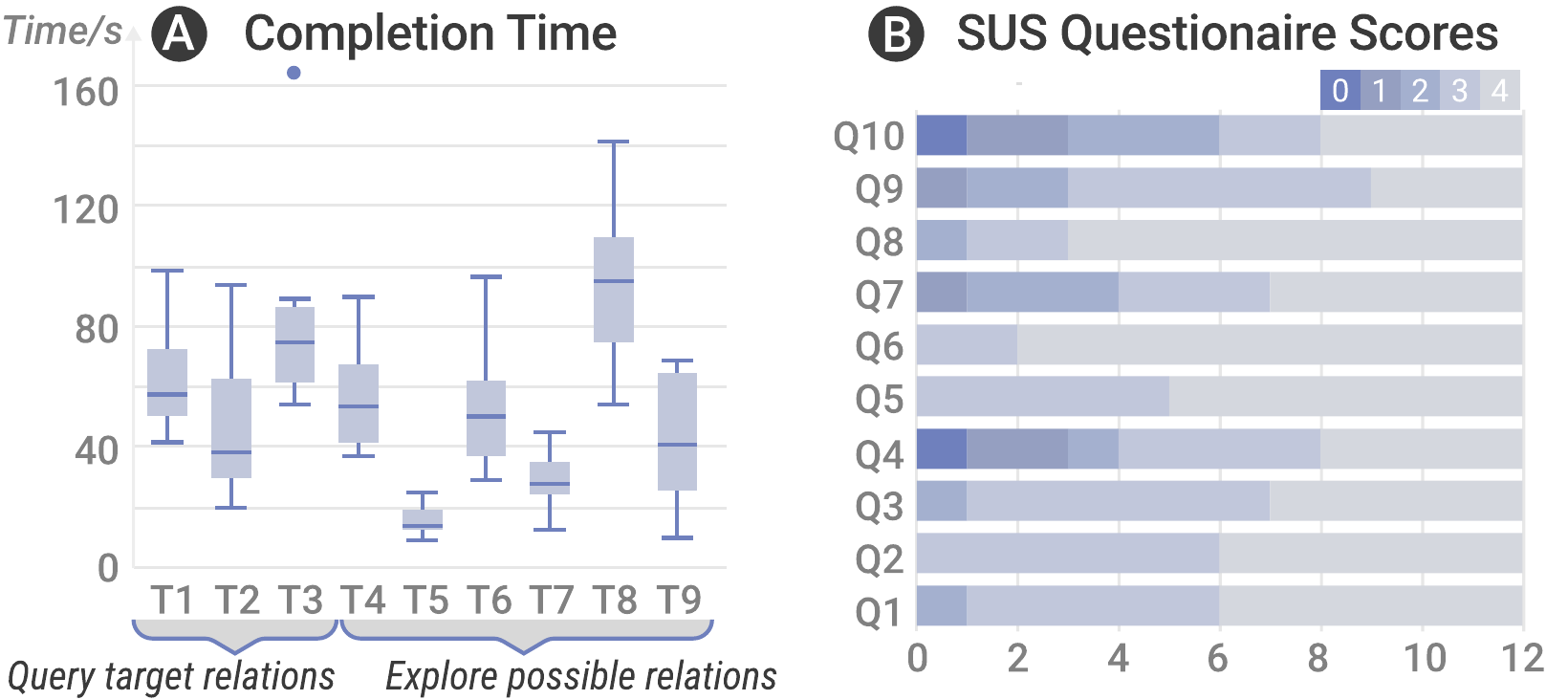}
  \caption{
The quantitative results of our task-based user study.
(A) The record of completion time reveals a high task pass rate and varied difficulty.
(B) We summarized the distribution of SUS scores given by subjects.
In general, subjects commented positively on the usability of RelaQ.
  }
  \label{fig8.sus}
\end{figure}

Most subjects passed tasks successfully, with an average passing rate for all tasks of 0.98.
There were two failures in total: S4 failed Q2 because he misunderstood the task description and deleted a relation constraint that should have existed;
S11 failed Q4 because he confused causality and correlation, resulting in adding a wrong relation.
For passed cases, the completion time is summarized in Fig.~\ref{fig8.sus}A.
The time for the subjects to complete a task ranged from 9 seconds to 163 seconds, with an average of 53 seconds.
Specifically, S4 spent more time in Q3 than others because he misunderstood the encoding of result distribution and browsed the results repeatedly on the matrix panel.
Overall, subjects took an average of 7.8 minutes to answer all tasks.

The subjects gave a SUS score of 81.04 on average, and a score above 80.3 is considered as the top 10\% of scores \cite{SUS}.
We also computed the detailed factors of SUS~\cite{susfactor}, with a usability score of 85.16 and a learnability score of 64.58.
Considering that the user study is conducted with non-experts, the learnability is still within an acceptable range.
Besides, based on expert interviews in the case study, we argue that the system is easy to learn and use for experts in the field.
Nonetheless, we plan to address the learning curve by providing more comprehensive tutorials and guidance in the future.
Generally, subjects highly evaluated the consistency of system design.
Most of them agreed that RelaQ is convenient to use and are willing to use it in the future. The details of the subjects' scores are shown in Fig.~\ref{fig8.sus}B, and we summarize their detailed feedback in the following section.

\subsection{Qualitative Usability}
We summarized subjects' feedback in terms of query, understanding, and guidance and improved RelaQ correspondingly.

\textbf{RelaQ enables easy formulation of query constraints (\textbf{R1}). } 
All subjects edited queries smoothly and confidently without confusion.
Specifically, S4 pointed out that sketches are consistent with his intent, enabling him to draw desired trends easily.
In terms of relations, S2 thought that the interaction of the relation query was very straightforward, as she could directly pick desired relations.
S10 stated that relation interactions require a low mental burden, even when specifying complex queries.
S7 appreciated smooth interactions: \textit{``I can check the changes of trends in real-time with animation while editing the query.''}
While subjects agreed with the intuitiveness, they also noted some limitations and suggested some improvements.
S4, S6, S9, and S11 commented that the interactions and visual encodings used in specifying queries cause a little memory burden.
S4 suggested hiding unnecessary buttons at the beginning and adding guided labels.

\textbf{RelaQ supports comprehensive understanding of query results (\textbf{R2}). }
Subjects found that RelaQ facilitated their understanding of query results.
S8 appreciated the node-link diagram's ability to show the abstract of the query and quickly convey the distribution of searched results: \textit{``It helps me quickly learn the distribution of searched results, such as whether a certain relation has matched a result.''}
S6 highly valued the combination feature for filtering subgraphs flexibly.
Some subjects, especially those (S3, 6-7, 11) who were familiar with native LineUp~\cite{Lineup}, raised their concern on the graph-based design.
S3 expressed that he was not confident about the results of auto combination: \textit{``When I was not so proficient in using RelaQ, I was not sure how the intermediate paths of node A and node B will be joined after combining them.''}
They commented that the graph-based design and auto combination were completely unfamiliar interactions for them, resulting in their lack of confidence at the beginning. 
However, they also agreed that it can be fully mastered after a few minutes of training. 
S6 commented: \textit{``The auto combination is very convenient. I can combine a subgraph by selecting the start and end nodes only.''}
S11 also stated that integrating graphs into LineUp is novel and useful in practical cases.

\textbf{Guidance effectively reduces the mental burden of exploration (\textbf{R3}). }
S5 highly appreciated recommendation: \textit{``Guidance effectively helped me find possible patterns, though I don't know much about domain knowledge.''}
Other subjects also gave positive comments, such as \textit{``effectively help narrow down search space''} (S9), and  \textit{``smart guidance''} (S12).
S10 expressed a strong preference for the guidance and successfully used it to query in a non-standard workflow, achieving the correct result. S10 was the only participant to use the guidance during the target-oriented query evaluation stage.
However, some limitations were noted. Specifically, S6 and S7 commented that the guidance should be automatically updated as queries are edited rather than requiring the user to click a corresponding button, which was perceived as inconvenient.

\section{Discussion}
In this section, we discuss the implications, limitations, and future work of RelaQ.

\textbf{Implications.}
This study systematically discusses the scope of relations between time series, taking the first step to explore the heterogeneous relation-driven approach of querying multiple time series.
RelaQ models the query problem as a graph-matching problem, not only concentrating on the sequential features of single time series but also on the structural features of multiple time series.
We propose a novel approach that combines a fuzzy query model and an interactive interface to support the direct and flexible specification of relations among time series without ambiguity, the in-depth understanding of queried temporal patterns, and reliable exploration of new queries.
Through a series of evaluations, RelaQ was verified as useful for solving real analysis problems in multiple domains.

\textbf{Limitations and future work.}
We collected valuable suggestions from users and optimized RelaQ.
There is still room for improvement in the long run, especially the trade-off between advanced functions and learnability and the concern about the scalability of the matching algorithm.

\underline{\textit{Support flexible relation computation.}}
Although we have chosen commonly used computations for each relation, experts prefer optional algorithms (EB) or even modifiable code blocks (EA) in extremely professional analysis tasks.
On one hand, RelaQ will become more powerful with the integration of modifiable computation.
On the other hand, such integration can increase the learning curve and mental burden for users.
Considering that RelaQ is our initial attempt at the relation-driven query of time series, we try to avoid excessive learning costs to encourage user adoption of this new query framework.
Thus, these advanced functions are left as future work.

\underline{\textit{Integrating retrieval with intelligent pattern mining.}}
The time complexity of the algorithm is nearly O($kN^2M$), where $k$ is the number of relations, $N$ is the number of variables, and $M$ is the length of time series.
A simple reduction of the complexity is discussed in Appendix 4.1.3.
This means the time costs highly depend on the data scale.
Unfortunately, when the scale of the time series reaches $10^8$, there is a noticeable lag in use.
We tested the time performance of RelaQ, and the results showed that it has good real-time performance in most cases. 
The average response time for queries of medium size and difficulty is 1.38s. 
More details can be found in Appendix 4.1.2.
Besides, RelaQ requires users to specify the match parameters: the box length and the sampling length, which means RelaQ cannot perform free-scale matching.
A proven effective approach to solving similar search problems is integrating intelligent pattern mining models.
Since RelaQ meets users' requirements in most daily analysis situations, we will enhance our model in the future by integrating intelligent approaches.
\section{Conclusion}
In this study, we conduct formative research to develop a semantic scope of time series relations and propose a novel approach for easy relation-driven time series queries through a prototype named RelaQ.
Evaluations include a quantitative user study and qualitative case study, demonstrating that RelaQ enables intuitive query and understanding of multiple time series with heterogeneous relations, allowing users to easily formulate complex queries through explicit relation interactions and sketches.
RelaQ is effective in real analysis tasks such as urban air pollution and EEG correlation analysis.
As an initial attempt at relation-driven time series queries, our future work will focus on optimizing RelaQ by integrating intelligent modules and adapting to more professional scenarios.

\section*{Acknowledgments}
The work was supported by National Key R\&D Program of China (2022YFE0137800), Key ``Pioneer'' R\&D Projects of Zhejiang Province (2023C01120), NSFC (U22A2032), the Collaborative Innovation Center of Artificial Intelligence by MOE and Zhejiang Provincial Government (ZJU), and Guangdong Basic and Applied Basic Research Foundation (2023B1515120078).

\bibliographystyle{IEEEtran}
\bibliography{checked}

\vspace{-1.2cm}
\begin{IEEEbiography}[{\includegraphics[width=1in,height=1.25in,clip,keepaspectratio]{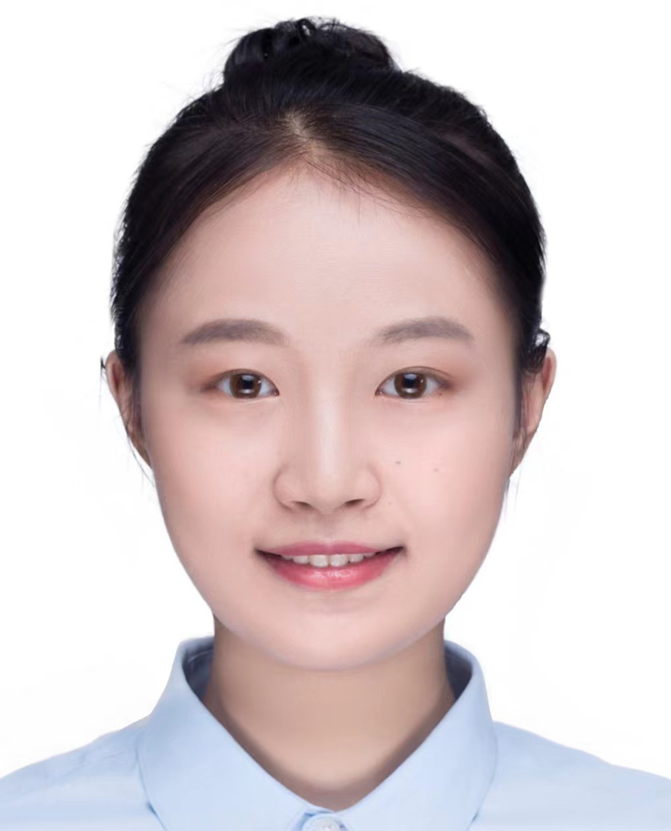}}]{Shuhan Liu} received her B.S. degree in computer science from Zhejiang University in 2021. She is currently pursuing a doctoral degree with the State Key Lab of CAD\&CG, Zhejiang University. Her research interests include spatiotemporal data mining, visualization, and industrial data visual analytics.
\end{IEEEbiography}
\vspace{-1.2cm}
\begin{IEEEbiography}[{\includegraphics[width=1in,height=1.25in,clip,keepaspectratio]{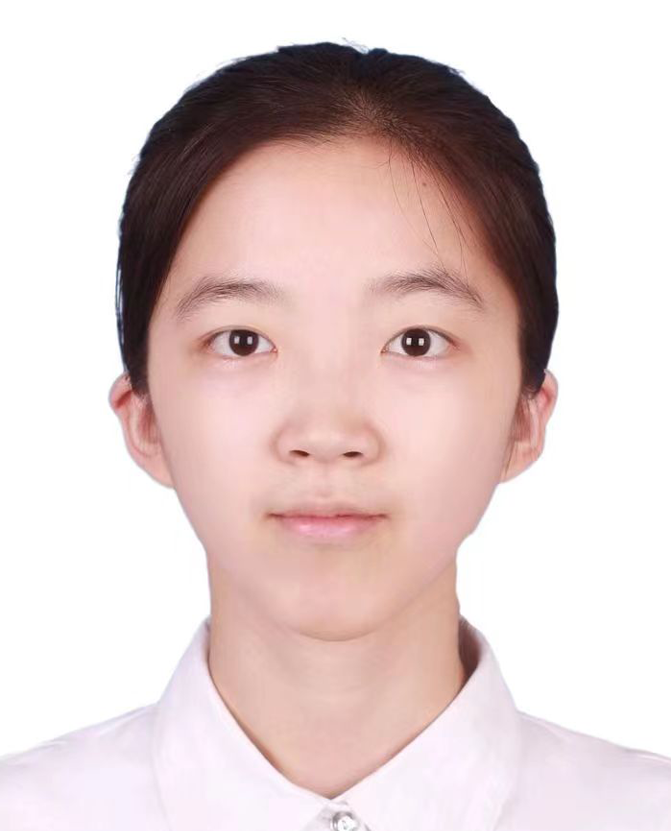}}]{Yuan Tian} received her B.S. degree in computer science from Zhejiang University in 2022. She is currently a Ph.D. student in the State Key Lab of CAD\&CG, Zhejiang University. Her research interests include machine learning for visualization and visual analytics. 
\end{IEEEbiography}
\vspace{-1.2cm}
\begin{IEEEbiography}[{\includegraphics[width=1in,height=1.25in,clip,keepaspectratio]{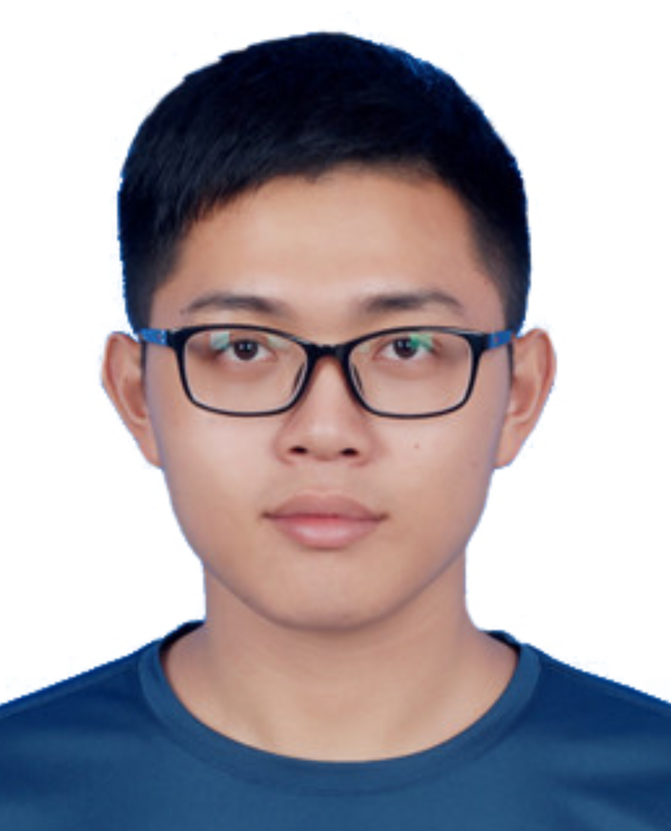}}]{Zikun Deng} is a tenure-track associate professor at School of Software Engineering, South China University of Technology. He received his Ph.D. degree in Computer Science from Zhejiang University in 2023. His research interests mainly include visual analytics, visualization, data mining, and their application in smart city, industry 4.0, and digital twins. For more information, please visit https://zkdeng.org.
\end{IEEEbiography}
\vspace{-1.2cm}
\begin{IEEEbiography}[{\includegraphics[width=1in,height=1.25in,clip,keepaspectratio]{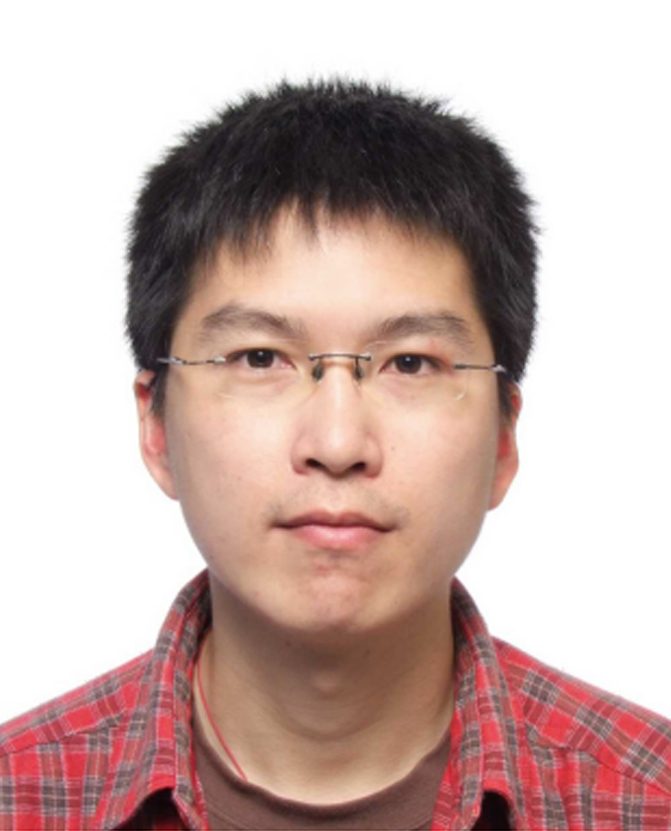}}]{Weiwei Cui} received the BS degree in computer science and technology from Tsinghua University, China, and the PhD degree in computer science and engineering from the Hong Kong University of Science and Technology, Hong Kong. He is a principal researcher at Microsoft. His primary research interest is visualization, with the focuses on democratizing visualization and AI-assisted design. For more information, please visit https://www.microsoft.com/en-us/research/people/weiweicu/.
\end{IEEEbiography}
\vspace{-1.2cm}
\begin{IEEEbiography}[{\includegraphics[width=1in,height=1.25in,clip,keepaspectratio]{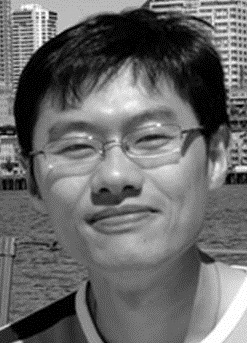}}]{Haidong Zhang} received the PhD degree in Computer Science from Peking University, China. He is a Principal Architect at Microsoft. His research interests include visualization and human-computer interaction.
\end{IEEEbiography}
\vspace{-1.2cm}
\begin{IEEEbiography}[{\includegraphics[width=1in,height=1.25in,clip,keepaspectratio]{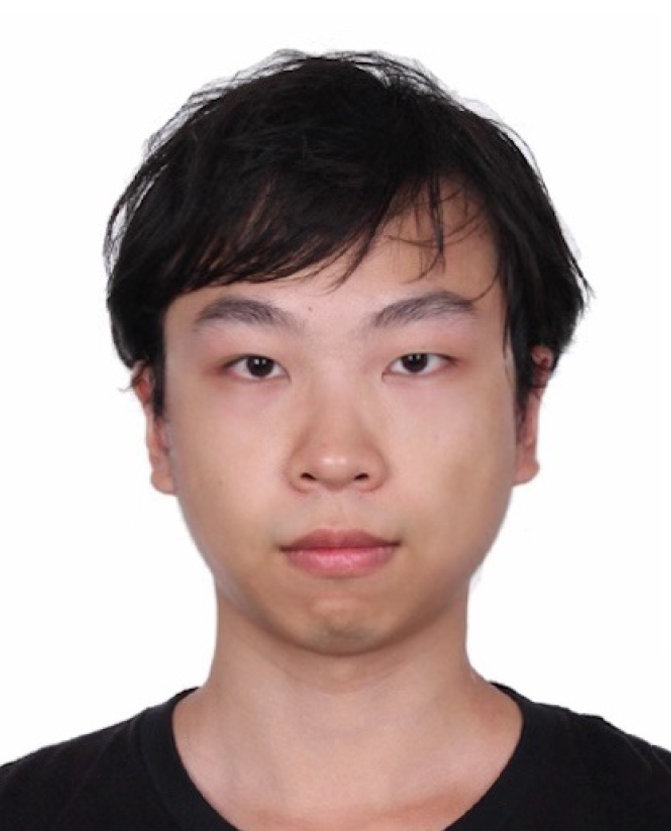}}]{Di Weng}  is a ZJU100 Young Professor at the School of Software Technology, Zhejiang University. His main research interest lies in information visualization and visual analytics, focusing on interactive data transformation and spatiotemporal data analysis. He received his Ph.D. degree in Computer Science from the State Key Lab of CAD\&CG, Zhejiang University. Before his current position, Dr. Weng was a researcher at Microsoft Research Asia from 2022 to 2023. For more information, please visit https://dwe.ng.
\end{IEEEbiography}
\vspace{-1.2cm}
\begin{IEEEbiography}[{\includegraphics[width=1in,height=1.25in,clip,keepaspectratio]{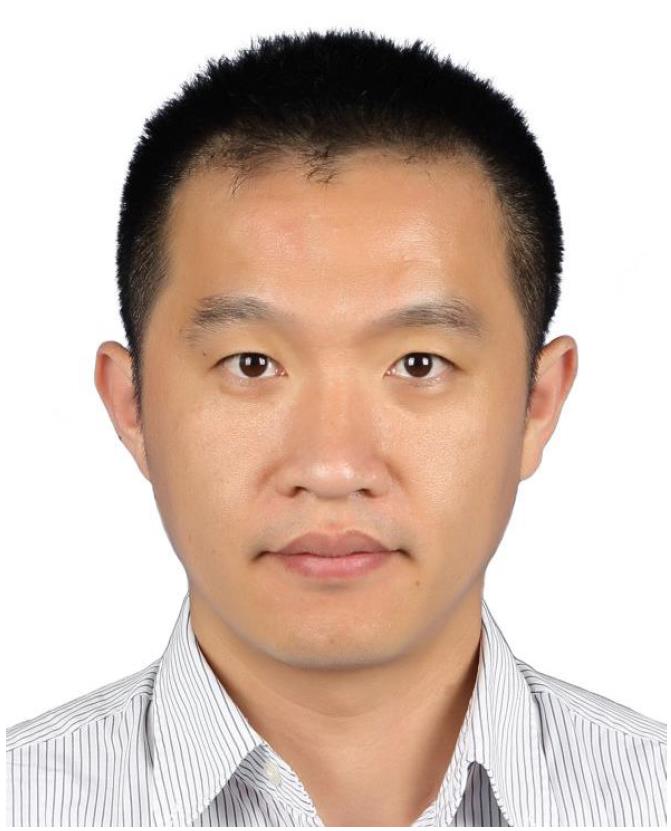}}]{Yingcai Wu} is a Professor at the State Key Lab of CAD\&CG, Zhejiang University.
His main research interests are information visualization and visual analytics, with focuses on urban computing, sports science, immersive visualization, and social media analysis. 
He received his Ph.D. degree in Computer Science from The Hong Kong University of Science and Technology. 
Prior to his current position, Dr. Wu was a postdoctoral researcher in the University of California, Davis, from 2010 to 2012, and a researcher at Microsoft Research Asia from 2012 to 2015. 
For more information, please visit http://www.ycwu.org.
\end{IEEEbiography}

\vfill

\end{document}